\documentclass[conference]{IEEEtran}
\IEEEoverridecommandlockouts

\usepackage{comment}
\usepackage{cite}
\usepackage{nohyperref}

\usepackage{fancyhdr}

\usepackage{multirow}
\usepackage{amsmath,amssymb,amsfonts}
\usepackage{algorithmic}
\usepackage{graphicx}
\usepackage{textcomp}
\usepackage{stmaryrd}
\usepackage{xcolor}
\def\BibTeX{{\rm B\kern-.05em{\sc i\kern-.025em b}\kern-.08em
    T\kern-.1667em\lower.7ex\hbox{E}\kern-.125emX}}
\newcommand{\red}[1]{\textcolor{red}{#1}}
\usepackage{amsmath,amsfonts,bm}

\def\eqref#1{equation~\ref{#1}}

\def\1{\bm{1}}

\def\mA{{\bm{A}}}
\def\mB{{\bm{B}}}

\def\mI{{\bm{I}}}

\def\mU{{\bm{U}}}
\def\mV{{\bm{V}}}
\def\mW{{\bm{W}}}
\def\mX{{\bm{X}}}

\DeclareMathAlphabet{\mathsfit}{\encodingdefault}{\sfdefault}{m}{sl}
\SetMathAlphabet{\mathsfit}{bold}{\encodingdefault}{\sfdefault}{bx}{n}

\usepackage{circledsteps}
\pgfkeys{/csteps/inner color=white}
\pgfkeys{/csteps/outer color=none}
\pgfkeys{/csteps/fill color=black}
\usepackage{pifont} 
\newcommand{\cmark}{\ding{51}}%
\newcommand{\xmark}{\ding{55}}
\definecolor{cadmiumgreen}{rgb}{0.0, 0.42, 0.24}
\usepackage{setspace}
\begin{document}
\title{NeuroUnlock: Unlocking the Architecture of Obfuscated Deep Neural Networks}

\author{
 \IEEEauthorblockN{Mahya~Morid~Ahmadi$^\dag$,
        Lilas~Alrahis$^\ddag$,
       Alessio~Colucci$^\dag$, 
       Ozgur~Sinanoglu$^\ddag$ and Muhammad~Shafique$^\ddag$}
  	\IEEEauthorblockA{$^\dag$\textit{Technische Universit\"at Wien (TU Wien), Vienna, Austria}\\
  	$^\ddag$\textit{Division of Engineering, New York University Abu Dhabi (NYUAD), Abu Dhabi, United Arab Emirates}\\
  	Email: \{mahya.ahmadi, alessio.colucci\}@tuwien.ac.at, \{lma387, ozgursin, muhammad.shafique\}@nyu.edu}\\ \vspace{-20pt}
 }

\maketitle

\renewcommand{\headrulewidth}{0.0pt}
\thispagestyle{fancy}
\lhead{}
\rhead{}
\chead{\copyright~2022 IEEE.
This is the author's version of the work.
The definitive Version of Record will be Published in the 2022 International Joint Conference on Neural Networks (IJCNN)}
\cfoot{}

%\setuptoappearheader{To appear at IJCNN 2022}
%\toappearheader

\begin{abstract}

The advancements of deep neural networks (DNNs) have led to their deployment in diverse settings, including safety and security-critical applications. As a result, the characteristics of these models (e.g., the architecture of layers and weight values/distributions) have become sensitive intellectual properties that require protection from malicious users. Extracting the architecture of a DNN through leaky side-channels (e.g., memory access) allows adversaries to (i) clone the model (i.e., build proxy models with similar accuracy profiles), and (ii) craft adversarial attacks. DNN obfuscation thwarts side-channel-based architecture stealing (SCAS) attacks by altering the run-time traces of a given DNN while preserving its functionality.

In this work, we expose the vulnerability of state-of-the-art DNN obfuscation methods (based on predictable and reversible modifications employed in a given DNN architecture) to these attacks. We present \textit{NeuroUnlock}, a novel SCAS attack against obfuscated DNNs. Our NeuroUnlock employs a sequence-to-sequence model that learns the obfuscation procedure and automatically reverts it, thereby recovering the original DNN architecture. We demonstrate the effectiveness of NeuroUnlock by recovering the architecture of $200$ randomly generated and obfuscated DNNs running on the Nvidia RTX 2080 TI graphics processing unit (GPU). Moreover, NeuroUnlock recovers the architecture of various other obfuscated (and publicly available) DNNs, such as the VGG-11, VGG-13, ResNet-20, and ResNet-32 networks. After recovering the architecture, NeuroUnlock automatically builds a near-equivalent DNN with only a $1.4\%$ drop in the testing accuracy. We further show that launching a subsequent adversarial attack on the recovered DNNs boosts the success rate of the adversarial attack by $51.7\%$ in average compared to launching it on the obfuscated versions.
Additionally, we propose a novel methodology for DNN obfuscation, \textit{ReDLock}, which eradicates the deterministic nature of the obfuscation and achieves $2.16\times$ more resilience to the NeuroUnlock attack.
We release the NeuroUnlock and the ReDLock as open-source frameworks\footnote{https://github.com/Mahya-Ahmadi/NeuroUnlock}.

\end{abstract}


\section{Introduction}
\label{Intro}
Deep learning has seen tremendous advances in the past few years, which has led to the employment of deep neural networks (DNNs) in diverse settings, from safety-critical applications like autonomous driving~\cite{kato2015open,tesla} 
to commercial applications such as image classification~\cite{krizhevsky2012imagenet}. The widespread usage of DNN models and their deployment on untrusted platforms (e.g., public clouds and edges) allows adversaries to access confidential model and data information. The sensitive characteristics of a DNN model include (i) the \textit{architecture} (i.e., the number, type, dimension, and connection topology of the layers), (ii) the \textit{parameters} (i.e., the weights, biases, etc.), and (iii) the \textit{hyper-parameters} used during training.

\begin{figure}[!t]
 \centering
 \includegraphics[width=\linewidth]{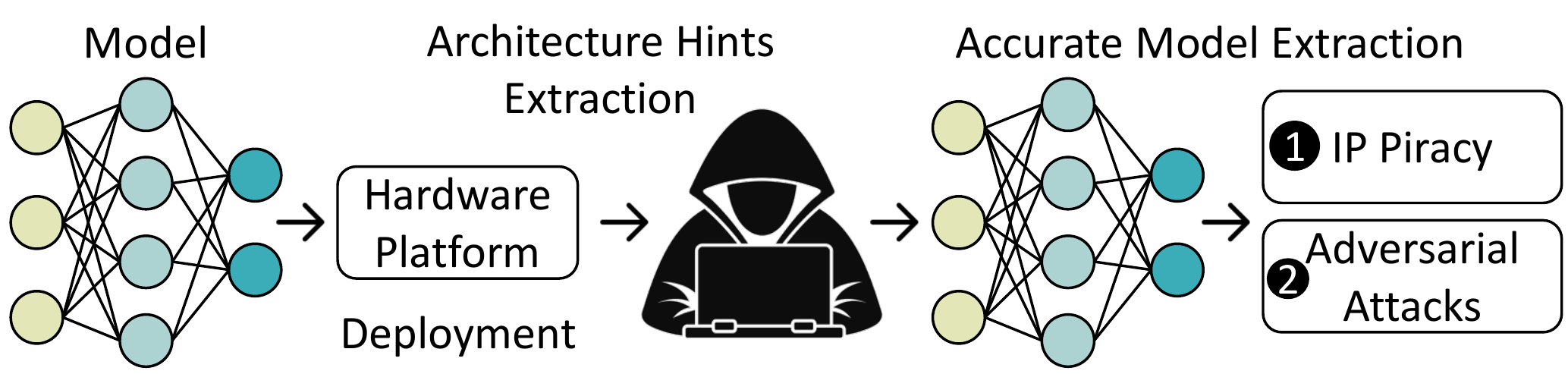}
 \vspace{-1.4em}
 \caption{DNN side-channel leakage facilitates piracy and adversarial attacks.}
 \vspace{-1em}
 \label{fig:example}
\end{figure}

Model extraction attacks extract \textit{(steal)} the model characteristics to counterfeit the intellectual property (IP) of a given DNN (i.e., build a model with a near-equivalent performance/predictions, see~\Circled{\scriptsize\textbf{1}} in Fig.~\ref{fig:example}) and may even facilitate (white-/grey-box) \textit{adversarial attacks} against the DNN system (see~\Circled{\scriptsize\textbf{2}} in Fig.~\ref{fig:example}), resulting in potentially catastrophic implications for the end-users and/or significant financial losses for IP providers/companies.\footnote{Adversarial attacks manipulate input samples to force the DNN to perform poorly on well-recognized outputs \cite{chakrabortySurveyAdversarialAttacks2021}.}
For example, X.~Hu~\textit{et al.} proposed \textit{DeepSniffer}~\cite{DeepSniffer} as a DNN side-channel-based architecture stealing (SCAS) attack that learns the correlation between the architecture hints (such as volumes of memory writes/reads) and the
DNN architecture. This work demonstrated that incorporating the architecture information almost triples the success rate of an adversarial attack.
\footnote{SCAS attacks require physical and system privilege access to the victim's hardware platform. Specifically, the DeepSniffer attack assumes access to the graphics processing unit (GPU) that encapsulates the victim DNN model~\cite{DeepSniffer}.}

\begin{table}[!t]
\centering
\vspace{-0.4em}
\caption{Comparison of the state-of-the-art methods that protect the DNN architecture from memory leakage SCAS attacks}
\vspace{-0.5em}
\label{tab:comparison}
\resizebox{0.5\textwidth}{!}{%
\setlength\tabcolsep{1pt} 
\renewcommand\arraystretch{1.1}
\begin{tabular}{cccc}
\hline
Defense & \begin{tabular}[c]{@{}c@{}}SCAS\\ Prevention\end{tabular} & \begin{tabular}[c]{@{}c@{}}Low Performance\\ Overhead\end{tabular} & \textbf{\begin{tabular}[c]{@{}c@{}}Resilience to\\NeuroUnlock Attack\end{tabular}} \\ \hline
Hardware-based Solutions~\cite{goldreich1996software,shi2011oblivious,karimi2020hardware} & \color{cadmiumgreen}{\cmark} & \red{\xmark} & \color{cadmiumgreen}{\cmark}\\ \hline
Memory Traffic Noise~\cite{DeepSniffer} & \red{\xmark} & \color{cadmiumgreen}{\cmark} & \red{\xmark} \\ \hline
NeurObfuscator~\cite{li2021neurobfuscator} & \color{cadmiumgreen}{\cmark} & \color{cadmiumgreen}{\cmark} & \red{\xmark} \\ \hline
\textbf{Proposed ReDLock} & \textbf{\color{cadmiumgreen}{\cmark}} & \textbf{\color{cadmiumgreen}{\cmark}} & \textbf{\color{cadmiumgreen}{\cmark}} \\ \hline
\end{tabular}%
}
\end{table}
\begin{figure*}[!t]
 \centering
 \includegraphics[width=\textwidth]{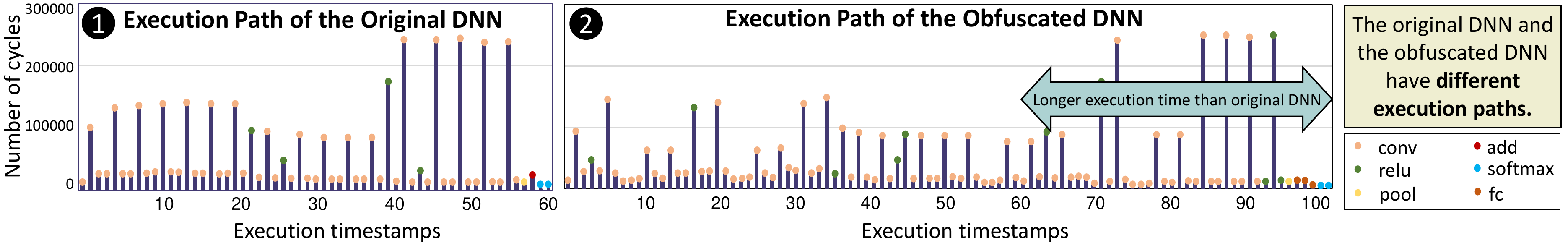}
 \vspace{-1.3em}
 \caption{DNN obfuscation alters the run-time traces (i.e., total run time, number of cycles, type of operations) while preserving the original functionality~\cite{li2021neurobfuscator}.}
 \vspace{-1em}
 \label{fig:altered}
\end{figure*}

To thwart SCAS attacks that exploit memory access information, prior works have proposed different methodologies, such as (i) preventing memory access leakage via hardware-based modifications~\cite{goldreich1996software,shi2011oblivious,karimi2020hardware}, (ii) introducing noise via fake memory traffic~\cite{DeepSniffer}, 
and (iii) hiding the DNN architecture via obfuscation at the scripting~(\textit{coding}) stage (e.g., using \textit{NeurObfuscator}~\cite{li2021neurobfuscator}).\footnote{We focus on memory, cache and timing leakage.} Among these different techniques, DNN obfuscation is a prominent one that incurs relatively low overhead. 
Methods (i) and (ii) suffer from several drawbacks, discussed next and summarized in Table~\ref{tab:comparison}.
\vspace{-1em}
\subsection{State-of-the-art and their Limitations}
\textbf{Latency Overhead:} The hardware-based oblivious random access memory (ORAM) schemes~\cite{goldreich1996software,shi2011oblivious} reduce the information leakage on the memory bus by encrypting (\textit{hiding}) the memory access patterns. However, the ORAM and other hardware-based architecture protection methods suffer from high performance overhead~\cite{hwsecurity}. For example, ORAM usage incurs $\approx \times10$ latency cost~\cite{liu2015ghostrider}.\footnote{Inference latency is the time required for a trained DNN to predict the output once a given input is provided.}

\textbf{Inefficiency:} Noisy memory traffic can be generated by introducing fake read/write operations. Unfortunately, SCAS attacks are robust to noise in architecture hints. For example, adding random noises within $5\%$ to $30\%$ of amplitude merely increases the average layer prediction error rate (LER) of the DeepSniffer attack from $0.08$ to $0.16$~\cite{DeepSniffer}. The LER should be close to or higher than $1$ to safely thwart SCAS attacks.
\footnote{The LER metric will be explained in the experimental section (Sec.~\ref{ExperimentalSetup}).}

\vspace{-0.5em}
\subsection{Key Research Challenges Targeted in this Work}
The state-of-the-art DNN obfuscation method, i.e., the NeurObfuscator~\cite{li2021neurobfuscator}, performs function-preserving operations (layer branching, layer deepening, etc.) during the scripting stage of a DNN, affecting the number of computations, latency, and the number of memory accesses, and hence, modifying the execution trace of the DNN and thwarting SCAS attacks. In the following, we discuss the major aspects of NeurObfuscator which make it resilient, and then we outline the key research challenges targeted in this work.
\begin{enumerate}
\item \textit{Altered execution trace:} Fig.~\ref{fig:altered} illustrates the execution paths of a randomly generated DNN (i.e., original DNN, see~\Circled{\scriptsize\textbf{1}}) and its obfuscated version (i.e., obfuscated DNN, see~\Circled{\scriptsize\textbf{2}}). It can be observed that the obfuscation changes the run-time trace, although both networks have the same functionality. As a result, SCAS attacks 
extract a wrong computation graph instead of the original one.\footnote{A computation graph is a graph representing the connections between the inputs, outputs, and parametric operations (e.g., convolution) of a single round of inference or a single iteration in the training stage of a DNN.} \textit{Thus, developing a method that reverts the obfuscation is required to recover the original architecture.}

\begin{figure}[!t]
 \centering
 \includegraphics[width=\linewidth]{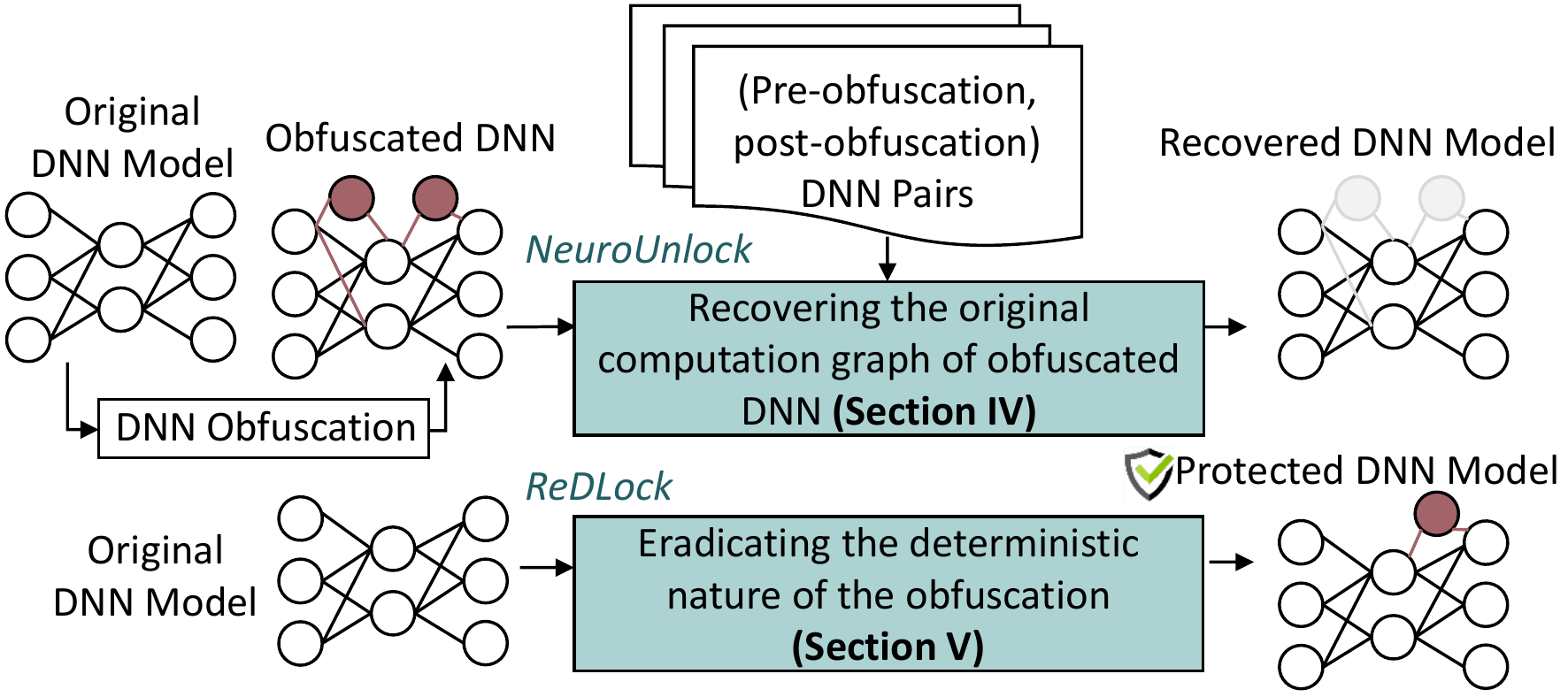}
 \vspace{-1.3em}
 \caption{The contributions presented in this paper are highlighted in blue boxes. }
 \vspace{-1em}
 \label{fig:Contribution}
\end{figure}
 
\item \textit{Guaranteed resilience:} The function-preserving mutations in NeurObfuscator are governed by an evolutionary algorithm (EA) that optimizes the DNN to maximize its resiliency against SCAS attacks. The evaluation stage of the EA launches an SCAS attack on the candidate obfuscated DNN and reports the LER error of the attack, which is considered as the fitness score of the obfuscated DNN. For instance, the NeurObfuscator achieves an LER $\approx3$ for an obfuscated ResNet-18 network with a latency cost of only $\times0.05$.

\end{enumerate}
\vspace{-0.5em}
\subsection{Our Novel Concept and Contributions}
We take the first step in evaluating the robustness of 
DNN obfuscation and show that the existing obfuscation exhibits deterministic characteristics that can be captured and undone using machine learning (ML) algorithms. We propose \textit{NeuroUnlock} platform that recovers the original DNN architecture before obfuscation, and thus, the design IP, through side-channel-based analysis guided by ML. Our novel contributions are summarized in Fig.~\ref{fig:Contribution}, and discussed below.
\begin{enumerate}
 \item \textbf{Recovering the original computation graph (Sec.~\ref{Methodology}):} NeuroUnlock captures the latency, memory access, and cache performance of an obfuscated DNN, and then translates the run-time traces to the obfuscated architecture (i.e., obfuscated sequence layer and corresponding dimensions) using an ML-based
approach. The obfuscated architecture is then reverted to its pre-obfuscation state using a sequence-to-sequence (Seq-2-Seq) model trained on (pre-obfuscation, post-obfuscation) DNN pairs from the training set.
 \item \textbf{Eradicating the deterministic nature of the obfuscation (Sec.~\ref{sec:proposed_obfuscation}):} 
 We propose reinforced DNN locking (\textit{ReDLock}) as a solution to thwart NeuroUnlock and other SCAS attacks. ReDLock randomly selects the number and type of obfuscation operations to be used. Therefore, ReDLock has no obfuscation patterns and consequently cannot be learned by ML algorithms.
\end{enumerate}

\textbf{Key Results:} Through our extensive experiments with a set of standard DNN models and $200$ randomly generated DNNs obfuscated using NeurObfuscator~\cite{li2021neurobfuscator} and deployed on the Nvidia RTX 2080 TI GPU platform, we demonstrate that NeuroUnlock attack is very effective with average LER reduction of $52\%$.
We further demonstrate that the proposed ReDLock obfuscation technique overcomes the limitations of the existing DNN obfuscation methods achieving $2.16\times$ more resilience against the NeuroUnlock attack in terms of higher LER introduced to the layer sequence.

\noindent \textbf{Open-Source Contribution:} The source code for the NeuroUnlock and ReDLock frameworks, along with the related datasets are accessible at \href{https://github.com/Mahya-Ahmadi/NeuroUnlock}{Github}.

Before proceeding to the technical sections, we present the background related to DNN obfuscation and SCAS attacks on DNNs, in Sec.~\ref{Background}.
\vspace{-0.3em}
\section{Background and Related Work}
\label{Background}
\vspace{-0.2em}
\subsection{Side-channel-based Architecture Stealing (SCAS) Attacks}
\begin{figure}[!t]
 \centering
 \includegraphics[width=\linewidth]{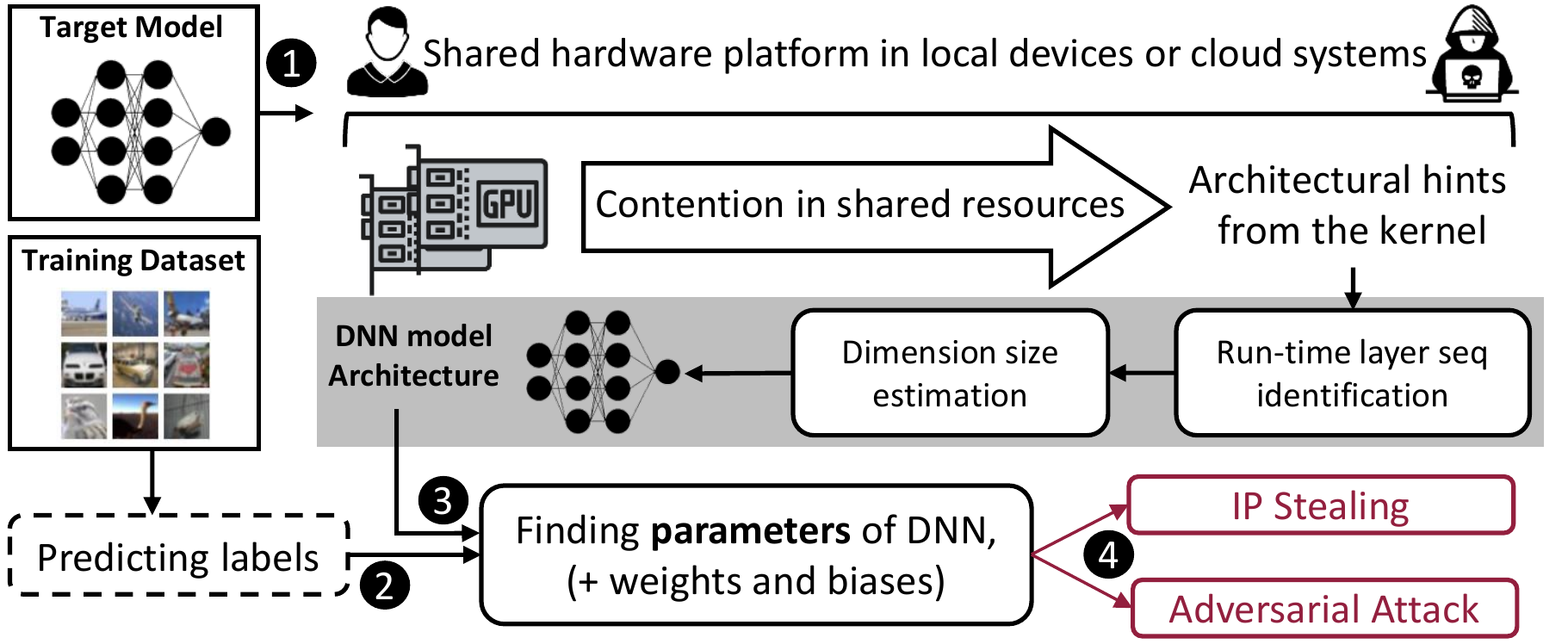}
 \vspace{-1em}
 \caption{Side-channel-based architecture stealing (SCAS) attack.
 }
 \vspace{-0.5em}
 \label{fig:ThreatModel}
\end{figure}

Model extraction attacks jeopardize the security of a DNN system by extracting confidential model characteristics such as the architecture and the parameters. Obtaining the exact characteristics of a DNN (i.e., extracting an identical DNN model) is practically challenging considering a black-box model~\cite{jagielski2020high} (i.e., accessing the interface of the victim model without knowledge of other DNN characteristics). Physical access to the DNN's hardware platform allows adversaries to leak confidential model characteristics through power, voltage, memory access, and timing side-channel information~\cite{rakin2021deepsteal} (see~\Circled{\scriptsize\textbf{1}} in Fig.~\ref{fig:ThreatModel}).
A substitute model is then built based on the extracted characteristics and later trained by querying the victim DNN or by using some publicly available labeled training dataset (see~\Circled{\scriptsize\textbf{2}} and \Circled{\scriptsize\textbf{3}} in Fig.~\ref{fig:ThreatModel}). 
The attacker can then leverage the substitute model to craft adversarial attacks against the DNN system (see~\Circled{\scriptsize\textbf{4}}). In this work, we focus on the attacks that exploit memory, cache, and timing-based leakage to expose the architecture of DNNs running on GPU devices.

\begin{figure}[!t]
 \centering
 \includegraphics[width=0.9\linewidth]{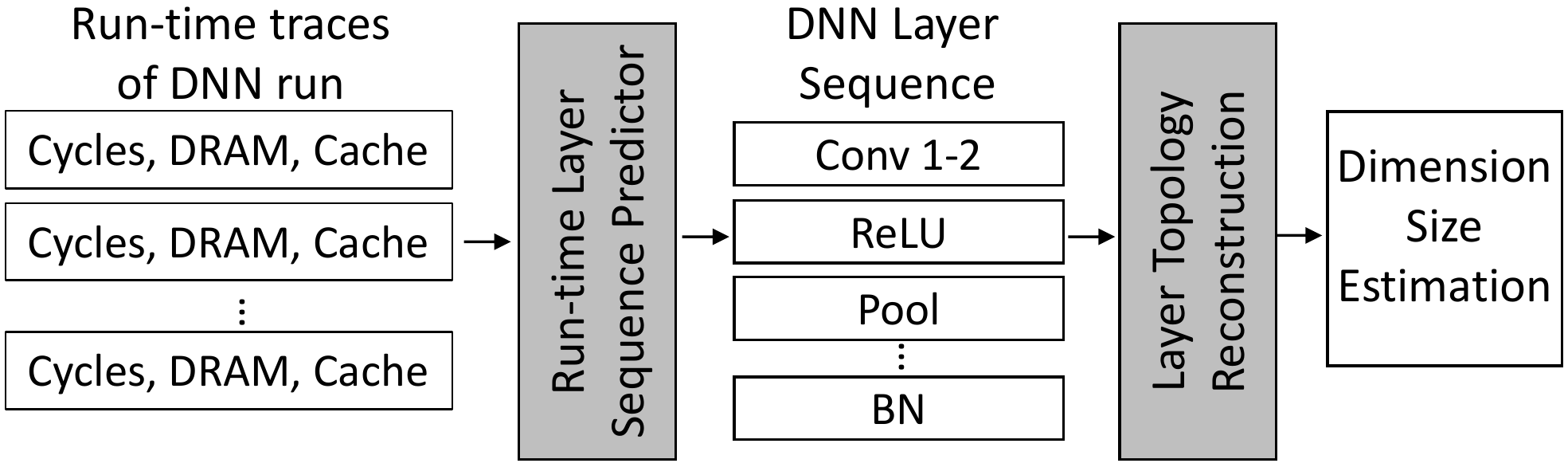}
 \vspace{-0.25em}
 \caption{Flow of the DeepSniffer attack~\cite{DeepSniffer}.
 }
 \vspace{-0.5em}
 \label{fig:DeepSniffer}
\end{figure}

In DeepSniffer~\cite{DeepSniffer} (shown in Fig.~\ref{fig:DeepSniffer}), a long-short-term-memory~(LSTM) model~\cite{lstm} infers the layer sequence of the victim DNN from its run-time trace, which is a time-series collection of different features such as latency, dynamic random access memory~(DRAM) access, and cache-hit rate.\footnote{The cache-hit rate represents the total number of cache hits divided by the number of cache requests.} Training the LSTM model requires profiling randomly generated DNNs on the target GPU. The layer sequence of each generated DNN gets encoded as a vector. Moreover, the run-time trace for each generated DNN gets extracted. A trace-sequence dataset is then built and used to train the LSTM.

After obtaining the LSTM predictions, the dimension for each layer is inferred based on its predicted operation and time-step position. Since the SCAS attacks rely on the run-time trace of the victim DNN to recover its architecture, DNN obfuscation aims to thwart such attacks by altering the execution trace using the function-preserving operations discussed in the next section.
\vspace{-0.3em}
\subsection{DNN Layer Sequence Obfuscation Operations~\cite{zhu2019eena,li2021neurobfuscator}}
\label{sec:DNN_Obfuscation}
\begin{figure}[!t]
 \centering
 \includegraphics[width=\linewidth]{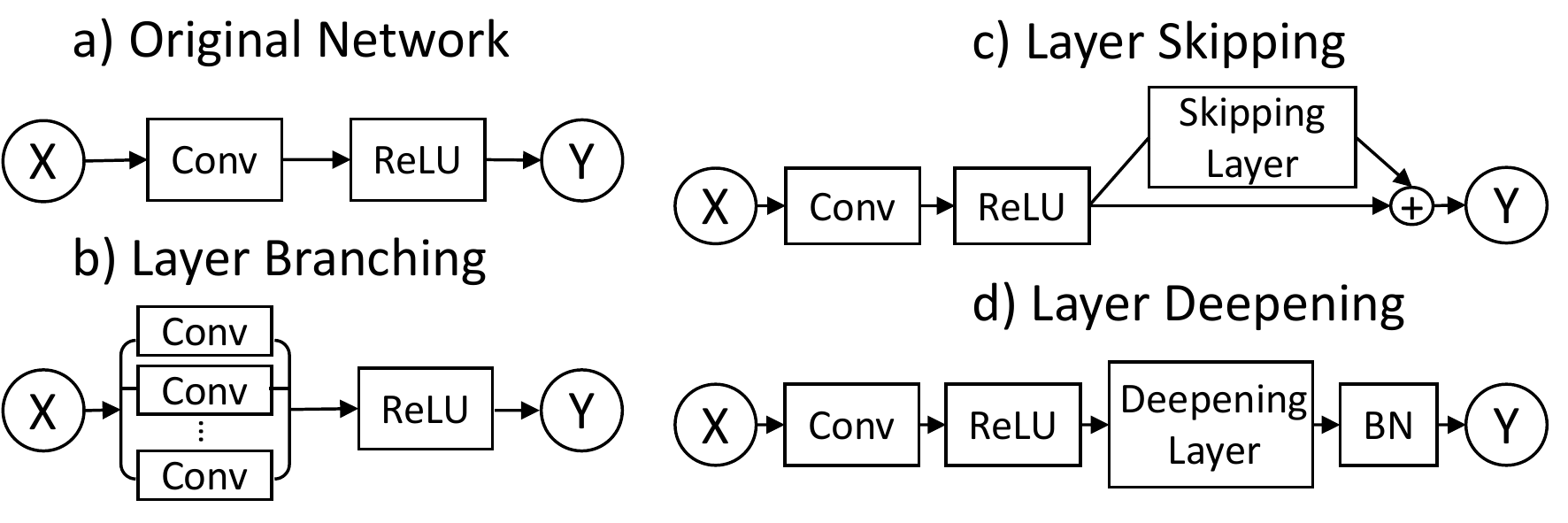}
 \vspace{-2em}
 \caption{Visualization of the different obfuscation operations~\cite{zhu2019eena, li2021neurobfuscator}.}
 \vspace{-1.2em}
 \label{fig:Obf_Operation}
\end{figure}

Let the matrix $\mW_{k1, k2, c, j}^{(i)}$ represent the $i^{th}$ convolutional layer to be modified. $k1$ and $k2$ represent the height and width of the convolution kernel, respectively, while $c$ and $j$ denote the input and output channel size, respectively. $\mX^{(i)}$ and $\sigma(\cdot)$ denote the input of the layer and the activation function (e.g., \textit{ReLU}), respectively. Fig.~\ref{fig:Obf_Operation}~(a) illustrates the original operator.

\begin{table}[!t]
\centering

\caption{The steps of an end-to-end attack against DNN systems}
 \vspace{-0.6em}
\label{tab:ThreatModel}
\setlength\tabcolsep{2pt} 
\renewcommand\arraystretch{1.1}
\small
\resizebox{0.48\textwidth}{!}{%
\setlength\tabcolsep{1pt} 
\renewcommand\arraystretch{1.1}
\begin{tabular}{ccccc}
\hline
\multirow{2}{*}{\begin{tabular}[c]{@{}c@{}}Required\\ Information\end{tabular}} & Step 1 & Step 2 & Step 3 & Step 4 \\ \cline{2-5} 
 & SCAS & \textbf{NeuroUnlock} & \begin{tabular}[c]{@{}c@{}}Model\\ Extraction\end{tabular} & \begin{tabular}[c]{@{}c@{}}Adversarial\\ Attack\end{tabular} \\ \hline
GPU Access & {\cmark} & {\cmark} & {\xmark} & {\xmark} \\ \hline
DNN Architecture & {\xmark} & {\xmark} & {\cmark} & {\xmark} \\ \hline
Obfuscation Tool & {\xmark} & {\cmark}& {\xmark} & {\xmark} \\ \hline
Train$/$Test Dataset & {\xmark} &{\xmark} & {\cmark} & {\cmark} \\ \hline
\end{tabular}%
}
\end{table}
\begin{figure*}[!t]
 \centering
 \includegraphics[width=\linewidth]{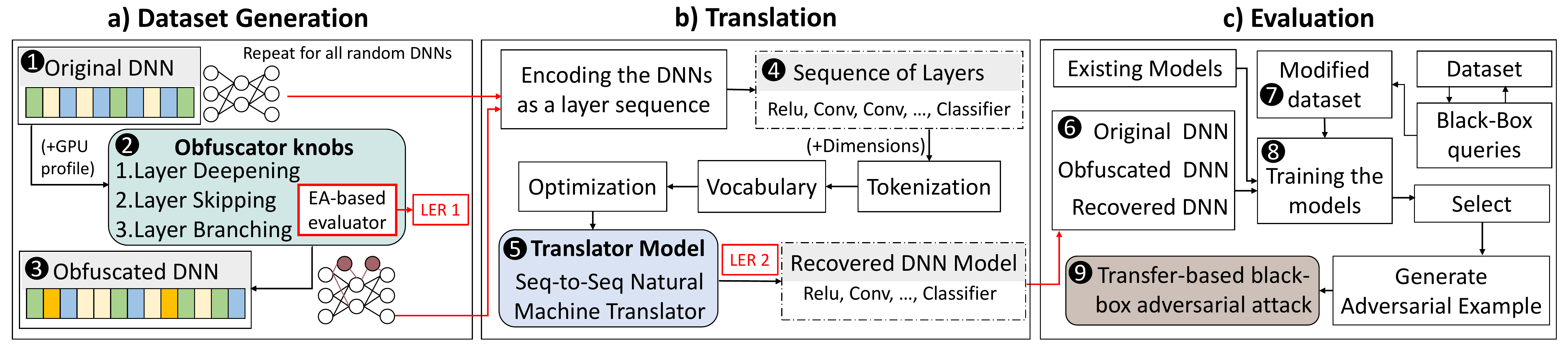}
 \vspace{-1em}
 \caption{Proposed NeuroUnlock methodology for unlocking the architecture of obfuscated DNNs.}
 \vspace{-0.5em}
 \label{fig:Methodology}
\end{figure*}
\textbf{Layer Branching:} splits a single layer operator into partial (smaller) operators, as illustrated in Fig.~\ref{fig:Obf_Operation}~(b). For example, a 2-D convolution layer (\textit{Conv2D}) $\mW_{k_1, k_2, c, j}^{(i)}$ can be deconstructed into two partial convolutions, as follows.
\begin{equation}
\begin{split}
\mU_{k_{1},k_{2},c,j/2}^{(i)} = \mW_{k_{1},k_{2},c,m}^{(i)} \quad m \in \left[0,\lfloor \frac{j}{2} \rfloor \right),\\
\mV_{k_{1},k_{2},c,j/2}^{(i)} = \mW_{k_{1},k_{2},c,m}^{(i)} \quad m \in \left[\lfloor \frac{j}{2} \rfloor,j \right)
\end{split}
\end{equation}
The final output is the concatenation of the two parts.
\begin{equation}
 \mU_{k_1, k_2, c, j/2}^{(i)} * \mX^{(i)} || \mV_{k_1, k_2, c, j/2}^{(i)} * \mX^{(i)}
\end{equation}
The splitting can also be performed in the input channel dimension.
The final output in this case is the addition of the two, as follows, where $\mX^{(i)}$ is sliced into $\mA^{(i)}$ and $\mB^{(i)}$.
\begin{equation}
 \mU_{k_1, k_2, c/2, j}^{(i)} * \mA^{(i)} + \mV_{k_1, k_2, c/2, j}^{(i)} * \mB^{(i)}
\end{equation}

\textbf{Layer Skipping:} inserts an extra Conv2D layer $\mU^{(i+1)}_{k_1, k_2, j, j}$ with all its parameters set to $0$ to maintain the functionality. An example is illustrated in Fig.~\ref{fig:Obf_Operation}~(c). The Conv2D layer will be reformulated as follows, where $\sigma \left( \mX^{(i+1)} \right)$ represents the activation output of the $i^{th}$ original layer.
\begin{equation}
\sigma \left( \mX^{(i+1)} \right) + \sigma \left( \mU^{(i+1)} * \mX^{(i+1)} \right) = \sigma \left( \mX^{(i+1)} \right) 
\end{equation} 

\textbf{Layer Deepening:} inserts an additional computational layer after the current layer activation and before \textit{batch normalization} (BN), as illustrated in Fig.~\ref{fig:Obf_Operation}~(d). If the preceding layer is linear, then the newly added $\mU^{(i+1)}$ is initialized as an identity matrix $\mI$ to maintain the functionality. Otherwise, $\mU^{(i+1)}_{k_1, k_2, j, j}$ can be generalized as:
\begin{equation}
\mU_{a,b,c,d}^{(i+1)}=
\left\{ 
 \begin{array}{lr}
 1 & a= \frac{k_{1}+1}{2} \wedge b= \frac{k_{2}+1}{2} \wedge c=d \\
 0 & \rm otherwise
 \end{array}
\right. 
\end{equation}
Layer deepening is valid as long as the activation function satisfies the following restriction, as the ReLU function does.
\begin{equation}
\forall x:\sigma(x)=\sigma \left( \mI*\sigma(x) \right)
\end{equation}

Post obfuscation, the computation graph is extracted and passed to the TVM\textsuperscript{\textregistered}~compiler~\cite{chen2018tvm}, which performs graph and operator-level optimizations.\footnote{DNN obfuscation can be categorized into sequence and dimension obfuscation. We focus on sequence obfuscation since the sequence identification stage is the most fundamental step in SCAS attacks.} In~\cite{li2021neurobfuscator}, the obfuscation operations are controlled by an EA platform that optimizes the latency cost and the resilience against SCAS attacks.

\vspace{-0.5em}
\section{Threat Model of Proposed NeuroUnlock}
We propose an attack on obfuscated DNNs that recovers the original DNN architectures before obfuscation.
In this section, we discuss the capabilities of the attacker and the different attack stages. The adversary has no prior knowledge of the victim DNN architecture, parameters, training algorithms, or hyperparameters. We focus on edge security, in which the attacker has (i)~system privilege access to the GPU platform encapsulating the victim DNN model, (ii)~the inputs and outputs (labels) of the model, and (iii) the DNN obfuscation tool to generate a dataset of obfuscated DNNs.
Consistent with most recent related works~\cite{dataset}, we further assume that the attacker has access to (iv)~a publicly available 
training dataset.

Table~\ref{tab:ThreatModel} lists all the steps of an end-to-end attack against an obfuscated DNN and the required information of each step.

\vspace{-0.3em}
\section{Proposed NeuroUnlock Attack}
\label{Methodology}
In this section, we provide an overview of the main steps of the NeuroUnlock attack (Fig.~\ref{fig:Methodology}), followed by detailed discussions of these steps.
\vspace{-0.3em}
\subsection{Dataset Generation}

\begin{figure}[!t]
 \centering
 \includegraphics[width=\linewidth]{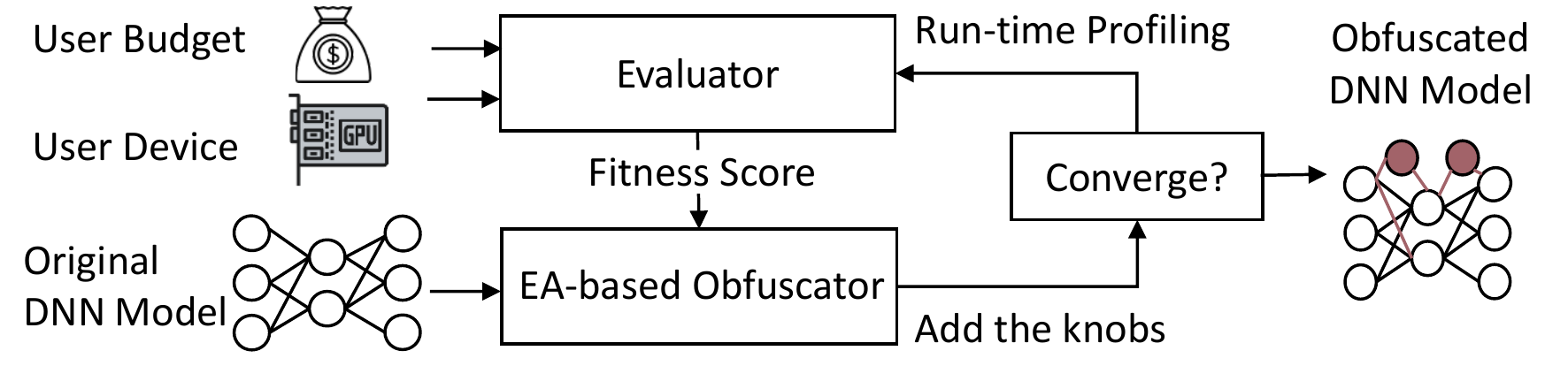}
 \vspace{-1.3em}
 \caption{Evolutionary algorithm (EA)-based DNN obfuscation~\cite{li2021neurobfuscator}.}
 \vspace{-1.5em}
 \label{fig:NeurObfuscator}
\end{figure}

 To launch NeuroUnlock against DNN obfuscation, we build a dataset of randomly-generated DNNs and their corresponding obfuscated versions, considering NeurObfuscator as a showcase, as illustrated in Fig.~\ref{fig:Methodology} (a). NeurObfuscator~\cite{li2021neurobfuscator} utilizes an EA platform to search for the best combination of sequence obfuscation knobs (operations) to obfuscate any DNN architecture. The EA platform considers the latency cost and the results of an SCAS attack against the obfuscated DNN as fitness scores (see Fig.~\ref{fig:NeurObfuscator}). Therefore, to obfuscate the DNNs, we need to train the SCAS attack model used in the evaluation stage of the obfuscation. Hence, the first stage of NeuroUnlock is building a dataset of random DNNs, profiling the GPU, collecting the corresponding run-time traces, training the SCAS attack model, and obfuscating the generated DNNs.
\begin{figure*}[!t]
 \centering
 \includegraphics[width=\linewidth]{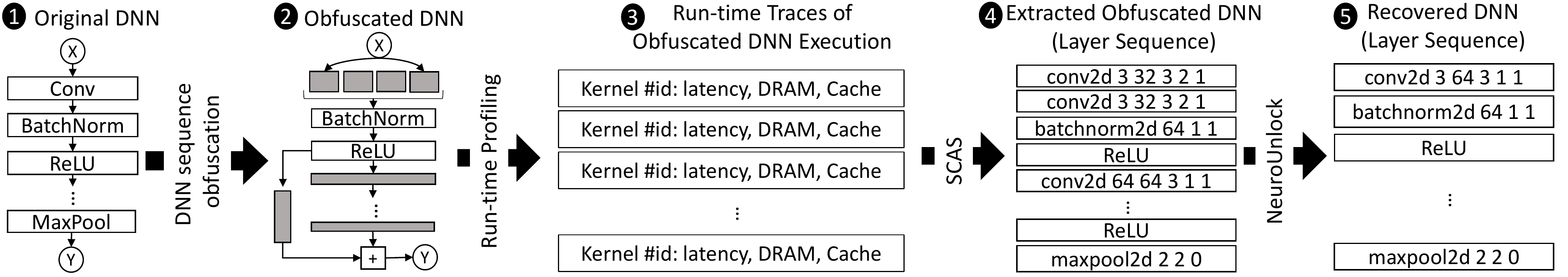}
 \vspace{-1em}
 \caption{The process of DNN obfuscation, extraction, and recovery by NeuroUnlock. }
 \vspace{-1em}
 \label{fig:Sequences}
\end{figure*}
\subsubsection{Random DNN Generation} 
\label{sec:random}
The number of computational layers for each DNN is picked as follows.

The number of Conv2D layers is randomly picked from the range of $[4 \shortrightarrow 12]$, the number of \textit{fully connected} (FC) layers is randomly picked from the range of $[1 \shortrightarrow 4]$. The Conv2D layers are then added with random output channel sizes in the range of $[16, 32, 64, 128, 256, 512, 1024]$. The output and the input dimensions of two stacked Conv2D layers are aligned. Some of the added Conv2D layers are picked (with a chance of $16\%$) and replaced with computing blocks from the \textit{ResNet} and \textit{MobileNet} networks. Following a binomial distribution, some of the remaining Conv2D layers are changed to \textit{Pooling} layers (i.e., max or average pool). After adding the Conv2D layers, the FC layers with dimensions in the range of $[16, 32, 64, 128, 256, 512]$ are stacked, followed by the final \textit{classification layer}
and the \textit{softmax} layer. A BN layer is added after each FC and Conv2D layer. \cite{lecunBackpropagationAppliedHandwritten1989}

We generate $5,000$ different DNNs (\textit{dataset~A}) with $3$ input channels each, width and height of $32$, and output size of $10$ (i.e., $10$ classes) to match the \textit{CIFAR-10} dataset setting (i.e., the input size of the first layer and the output size of the last layer are the same for all the generated DNNs). The SCAS sequence predictor (evaluator) is trained using this set of DNNs (Sec.~\ref{sec:SCAS_training}).
Another set of $2,000$ different DNNs (\textit{dataset~B}) is generated to be obfuscated as the training dataset for NeuroUnlock~(see~\Circled{\scriptsize\textbf{1}}~$\shortrightarrow$~\Circled{\scriptsize\textbf{3}} in Fig.~\ref{fig:Methodology}).
We further obfuscate a set of typical DNN models (\textit{dataset~C}) such as VGG-11, VGG-13, ResNet-20, and ResNet-32 networks trained on the CIFAR-10 dataset to evaluate the performance of NeuroUnlock. We further launch a subsequent adversarial attack on the VGG-11 DNN (Sec.~\ref{sec:adversarial}).
\subsubsection{SCAS Sequence Predictor}
\label{sec:SCAS_training}

The DNNs in \textit{dataset~A} are used to profile the target GPU. The run-time traces of the networks are collected using the Nsight\textsuperscript{\textregistered}~Compute profiler~\cite{Nsight} and used to train the sequence predictor. 
Nsight returns accurate trace information, which includes the number of cycles (i.e., latency), DRAM access and cache performance for each issued operator of the DNN, executing in inference mode. 
The employed SCAS sequence prediction model is an average ensemble of $5$ LSTM networks with hidden dimensions of $64$, $96$, $128$, $256$, and $512$, respectively. Each predictor consists of a single LSTM layer with a connectionist temporal classification decoder, 
as used in DeepSniffer~\cite{DeepSniffer,li2021neurobfuscator}. After constructing the layer sequence, the dimensions of the layers are estimated. For this, DeepSniffer first finds the feature map size of the ReLU layers using the memory traffic volume and then propagates the sizes since the input size of each layer is equal to the output size of the previous one. The accuracy of this estimation is platform-dependent and is around $98\%$. Hence, in our work we focus on recovering the layer sequence and assume $100\%$ correct dimension recovery similar to the work in~\cite{li2021neurobfuscator}.
\begin{figure}[!t]
 \centering
 \includegraphics[width=\linewidth]{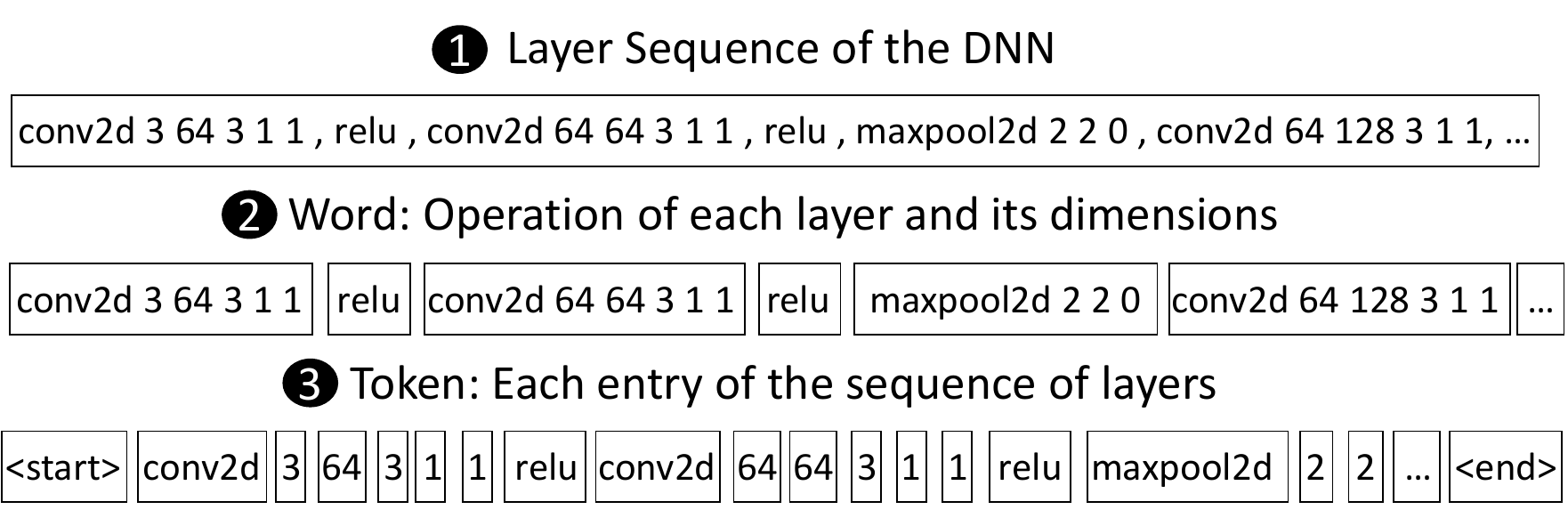}
 \vspace{-1.3em}
 \caption{NeuroUnlock layer sequence encoding and terminology.
 }
 \vspace{-1em}
 \label{fig:Dataset}
\end{figure}

\subsection{Obfuscated DNN Translation}
The DNNs in \textit{dataset~B} are first obfuscated (see~\Circled{\scriptsize\textbf{1}} and~\Circled{\scriptsize\textbf{2}} in Fig.~\ref{fig:Sequences}), and then, their run-time traces on the target GPU are extracted (see~\Circled{\scriptsize\textbf{3}} in Fig.~\ref{fig:Sequences}). The SCAS attack stage recovers an obfuscated layer sequence for each obfuscated DNN (see~\Circled{\scriptsize\textbf{4}} in Fig.~\ref{fig:Sequences}). NeuroUnlock takes the SCAS recovered obfuscated layer sequence, feeds it to a neural machine translation (NMT) model (i.e., seq-2-seq), and recovers the original layer sequence (see~\Circled{\scriptsize\textbf{5}} in Fig.~\ref{fig:Sequences}). A dataset of the SCAS extracted obfuscated DNNs and their corresponding original versions are used to train the NMT model employed in NeuroUnlock. The steps followed to build the translator and unlock the obfuscated DNNs are discussed next and summarized in Fig.~\ref{fig:Methodology}~(b).

\begin{figure}[!t]
 \centering
 \includegraphics[width=\linewidth]{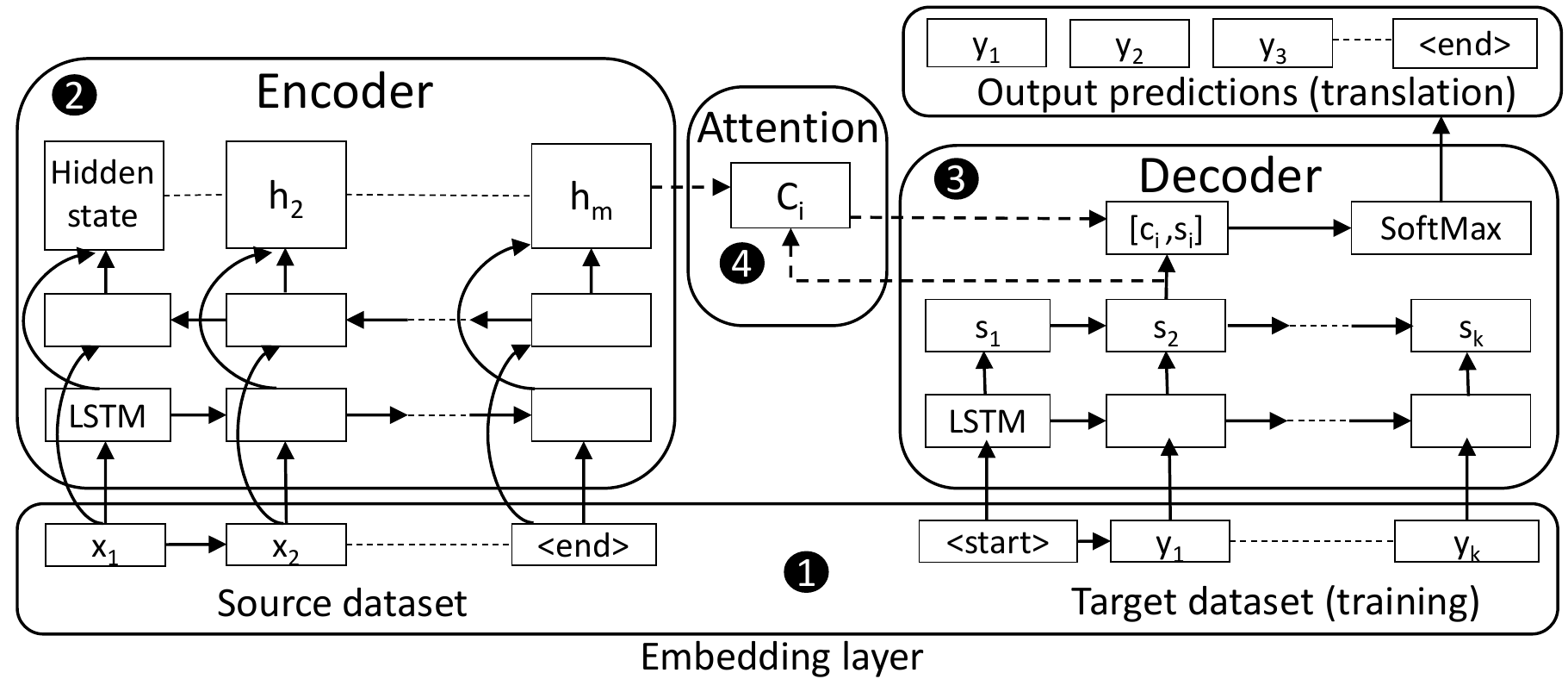}
 \vspace{-1.3em}
 \caption{Adopted natural machine translation (NMT) architecture~\cite{OpenNMT}.}
 \vspace{-1em}
 \label{fig:OpenNMT}
\end{figure}
\subsubsection{Translation Process} NMT models are commonly used in automated translation~\cite{OpenNMT} and, for our work, the goal is to translate the post-obfsucation (source) to the pre-obfuscation (target) layer sequence (see~\Circled{\scriptsize\textbf{5}} in Fig.~\ref{fig:Methodology}). First, we define the terms used in the adopted NMT approach (also summarized in Fig.~\ref{fig:Dataset}). The layer sequence of each DNN (both the original and the obfuscated) is first encoded in vector form (see~\Circled{\scriptsize\textbf{1}} in Fig.~\ref{fig:Dataset}). The parameters of each layer (i.e., ``operator type, input channel size, output channel size, kernel size, stride, and padding'') are preserved and included in the encoding as a single \textit{word} (see~\Circled{\scriptsize\textbf{2}} in Fig.~\ref{fig:Dataset}). For example, the encoding for the first Conv2D layer in \Circled{\scriptsize\textbf{1}} in Fig.~\ref{fig:Dataset} ``conv2d $3$ $64$ $3$ $1$ $1$'' tells that it is a Conv2D operation with input size of $3$, output size of $64$, $3\times3$ kernel size, stride of $1\times1$, and padding of $1\times1$. Each layer sequence is represented by words separated by commas. Each word is composed of different \textit{tokens} split by spaces (see~\Circled{\scriptsize\textbf{3}} in Fig.~\ref{fig:Dataset}). The unique tokens $<start>$ and $<end>$ are added to the start and end of each sequence. A \textit{vocabulary} is the set of all unique tokens along with their occurrence counts in the entire dataset. The adopted NMT model has four main components; (i) pre-processing, (ii) encoder, (iii) attention, and (iv) decoder (see Fig.~\ref{fig:OpenNMT}). Next, we discuss each component in detail.

The \textbf{pre-processing and embedding} step extracts the vocabulary from the dataset~(\Circled{\scriptsize\textbf{1}} in Fig.~\ref{fig:OpenNMT}) and prepares the input sequences by converting each token into its assigned index in vocabulary. During inference, the source sequences are passed to the encoder, while empty tokens $<>$ are passed to the decoder to enable the translation. An embedding layer converts the sequence of input tokens, i.e., layer operator and dimension information, into a floating-point vector which can be used as an input to the encoder. We construct an embedding layer with an output dimension of $78$ and an input size of at most $500$ tokens
 per sequence. This input size is enough to represent any DNN in our datasets.

The \textbf{encoder} (see \Circled{\scriptsize\textbf{2}} in Fig.~\ref{fig:OpenNMT}) processes each sequence after it has been transformed by the embedding layer. The encoder consists of a 2-layer LSTM network with a hidden state dimension of $500$.
The output of LSTM network updates the hidden states of the encoder
based on the input tokens, which will be used by the attention and the decoder to generate the output sequence.

The \textbf{decoder}~(see \Circled{\scriptsize\textbf{3}} in Fig.~\ref{fig:OpenNMT}) uses the hidden state updated by the encoder to produce the output sequence in vector form. This output needs to be converted back into a proper layer sequence by inverting the operation done by the embedding layer, i.e. generating the tokens from the floating-point vectors. The decoder is also a 2-layer LSTM network with a hidden dimension of $500$.

The \textbf{attention} component constructs a context vector that assists the decoder in processing the data, especially useful for sequences longer than encoder dimensions.
During the training phase, attention takes the state of the decoder network and combines it with the encoder hidden states~(see \Circled{\scriptsize\textbf{4}} in Fig.~\ref{fig:OpenNMT})
and feeds (forward) it to the decoder for the next prediction. Attention uses matrix multiplications to compute a weighted average of the encoder states to transform the internal state of decoder.
 During NMT model training, 
each output sequence is compared to the target sequence using the negative log likelihood loss (NLLLoss) function, and the error is backpropagated to update the parameters of the model.

\begin{figure}[!t]
 \centering
 \includegraphics[width=\linewidth]{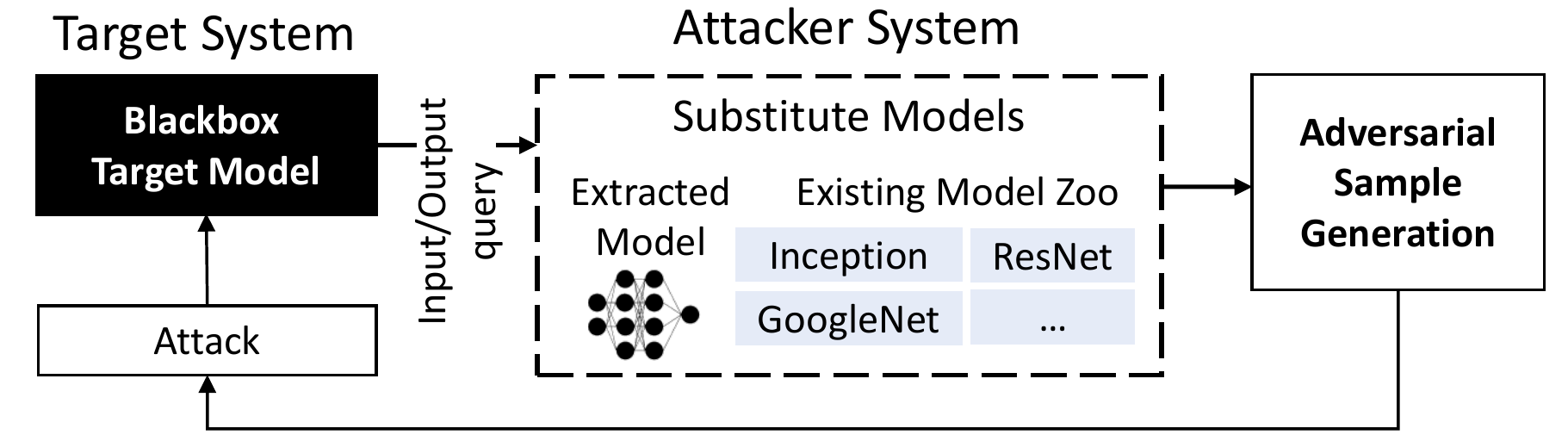}
 \vspace{-1em}
 \caption{Adopted transfer-based adversarial attack on DNNs.}
 \vspace{-1em}
 \label{fig:Adversarial}
\end{figure}

\vspace{-0.2em}
\subsection{Extracted DNN Evaluation}
\label{sec:adversarial}
To evaluate our methodology (as shown in Fig.~\ref{fig:Methodology}~(c)), we launch an adversarial attack against the recovered DNNs by NeuroUnlock (see~\Circled{\scriptsize\textbf{6}} in Fig.~\ref{fig:Methodology}). The goal is to show that the success rate of the adversarial attack increases after incorporating the architecture information recovered by NeuroUnlock.
The flow of the adopted transfer-based adversarial attack~\cite{AdvAtt} is illustrated in Fig.~\ref{fig:Adversarial}. The attack has three fundamental stages; (i) building/training a substitute model, (ii) creating adversarial samples, (iii) applying the adversarial samples. 
Substitute models for built for the target VGG-11 DNN model (the methodology is applicable to other DNN models as well). We build two models, the first one is built using the obfuscated DNN architecture, and the second one is built using the architecture information recovered by NeuroUnlock. Each substitute model is trained using input/output queries of the target DNN (see~\Circled{\scriptsize\textbf{6}}~$\shortrightarrow$~\Circled{\scriptsize\textbf{8}} in Fig.~\ref{fig:Methodology}). Next, the adversarial samples are created by adding small noises to the input sample that are not practically visible, but change the output of the classification. 
 We use a gradient-based approach to choose the location and the amount of added noise.
Then, we employ an ensemble-based method to boost the attacking success rate, based on the hypothesis that if an adversarial sample  remains adversarial for several models, it is more likely to be effective against the black-box model as well. These models are ResNet-34, InceptionV3, and GoogleNet.
The created adversarial samples are then used to attack the victim DNN model (see~\Circled{\scriptsize\textbf{9}} in Fig.~\ref{fig:Methodology}). This attack flow is the same as the one used in the DeepSniffer adversarial attack~\cite{DeepSniffer}.

\vspace{-0.3em}
\section{Proposed ReDLock Solution}
\label{sec:proposed_obfuscation}
To further support the intuition of NeuroUnlock, we propose ReDLock as a random DNN obfuscation method. The goal is to demonstrate that if the DNN obfuscation procedure is randomized (i.e., with no clear obfuscation pattern), it will be difficult for NeuroUnlock and other SCAS attacks to undo the obfuscation and recover the original DNN architecture.

\begin{figure}[!t]
     \centering
     \includegraphics[width=\linewidth]{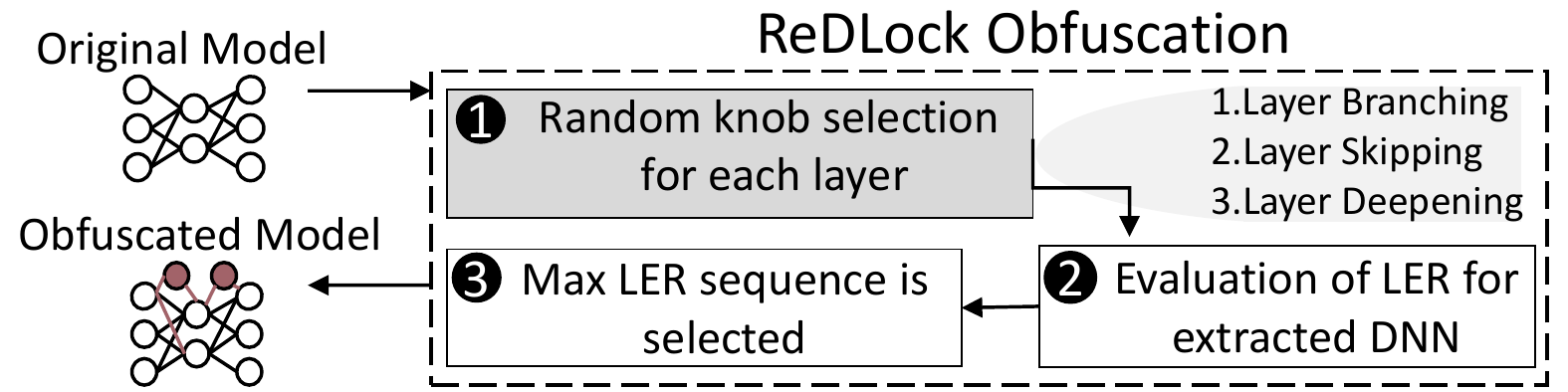}
     \vspace{-1em}
     \caption{The proposed ReDLock DNN obfuscation.
     }
     \vspace{-1em}
     \label{fig:ReDLock}
\end{figure}

The flow of the proposed ReDLock obfuscation is illustrated in Fig.~\ref{fig:ReDLock}. We employ all the sequence obfuscation operations discussed in Sec.~\ref{sec:DNN_Obfuscation}, i.e., layer branching, skipping, and deepening (\Circled{\scriptsize\textbf{1}} in Fig.~\ref{fig:ReDLock}).
For each computational layer of a given DNN, the obfuscation operations are randomly enabled and configured, as follows.
For each considered obfuscation operation, a choice is randomly made (with $50\%$ probability) as to whether to use the obfuscation operation or not. Hence, a layer could be obfuscated using 1 (with $37.5\%$ probability), 2 (with $37.5\%$ probability), 3 obfuscation operations (with $12.5\%$ probability), or none (with $12.5\%$ probability). We perform ReDLock obfuscation for $20$ times on the original model, obtaining $20$ different obfuscated versions of the same DNN. An SCAS attack is launched on all the obfuscated versions. The SCAS attack reports the LER between the extracted obfuscated DNN and the original DNN (\Circled{\scriptsize\textbf{2}} in Fig.~\ref{fig:ReDLock}). The obfuscated version with the highest LER is selected as the final obfuscated DNN (\Circled{\scriptsize\textbf{3}} in Fig.~\ref{fig:ReDLock}).

\section{Experimental Setup}
\label{ExperimentalSetup}
\begin{figure}[!t]
 \centering
 \includegraphics[width=\linewidth]{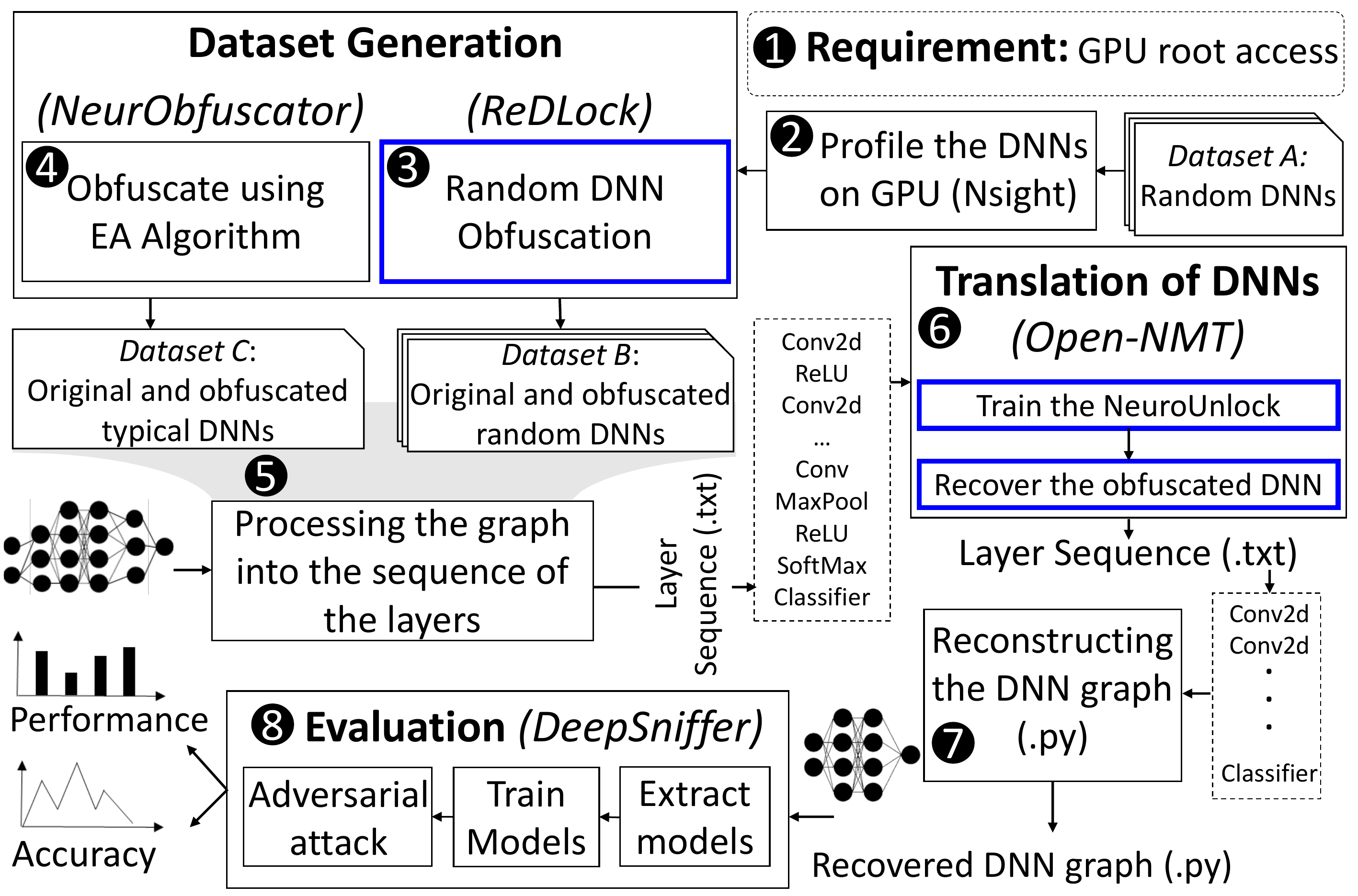}
 \vspace{-1.5em}
 \caption{The tool-flow and experimental setup of NeuroUnlock and ReDLock (highlighted in blue).
 }
 \vspace{-1em}
 \label{fig:Toolflow}
\end{figure}

In this section, we describe the tool-flow and the experimental setup adopted for evaluating the NeuroUnlock and ReDLock frameworks (summarized in Fig.~\ref{fig:Toolflow}).

\textbf{GPU Platform:} We chose the Nvidia RTX 2080 Ti GPU as a showcase. However, our methods are generally applicable to other GPU platforms. Nsight\textsuperscript{\textregistered}~Compute~\cite{Nsight} is used for profiling the GPU, which requires privileged access to the performance counters (\Circled{\scriptsize\textbf{1}} in Fig.~\ref{fig:Toolflow}). We generate a dataset of random DNNs (\textit{dataset~A}) as discussed in Sec.~\ref{sec:random} to profile them on the GPU (\Circled{\scriptsize\textbf{2}} in Fig.~\ref{fig:Toolflow}).

\textbf{DNN Obfuscation:} We evaluate NeuroUnlock on DNNs (\textit{dataset~B} and \textit{dataset~C}) obfuscated using our proposed ReDLock obfuscation and NeurObfuscator~\cite{li2021neurobfuscator} (\Circled{\scriptsize\textbf{3}} and \Circled{\scriptsize\textbf{4}} in Fig.~\ref{fig:Toolflow}). $80\%$ of the DNNs in \textit{dataset~B} are used to train the NMT translator of NeuroUnlock (\Circled{\scriptsize\textbf{6}} of Fig.~\ref{fig:Toolflow}). The remaining $20\%$ of the DNNs in \textit{dataset~B} and the full \textit{dataset~C} are used to evaluate the performance of the attack.
The computation graphs of the obfuscated/original DNNs are stored in Python files, which are then processed to obtain the corresponding layer sequences in text form (\Circled{\scriptsize\textbf{5}} in Fig.~\ref{fig:Toolflow}).

\textbf{NMT-based Translation:} The DNNs are organized into a dataset of post-obfuscation (source) and pre-obfuscation (target) layer sequences.
To build the NMT model, we employ \textit{Open-NMT}~\cite{OpenNMT}, an open-source project built on the \textit{PyTorch} deep learning framework~\cite{pytorch}. We train the NMT model for $25$ epochs, using the 
stochastic gradient descent optimization implementation~\cite{sgd}, called the \textit{SGD} optimizer in PyTorch, with a learning rate of $1$. After training, the NMT translator model predicts a recovered DNN architecture (i.e., layer sequence and dimensions) given an obfuscated DNN.

The computation graphs of the recovered DNNs in \textit{dataset C} are extracted from the recovered layer sequence, and described as Python model files (\Circled{\scriptsize\textbf{7}} in Fig.~\ref{fig:Toolflow}). Each extracted model will be trained to launch a subsequent adversarial attack, as discussed in Sec.~\ref{sec:adversarial}. To this end, all the models are trained on the outputs obtained by processing the CIFAR-10 dataset~\cite{cifar10Krizhevsky2009LearningML} using the original DNN (i.e., black-box model queries). For the optimization process, we employ the \textit{Adam}~\cite{DBLP:journals/corr/KingmaB14} SGD optimizer with a learning rate of $0.001$. The models are trained for $30$ epochs, and the model with the highest validation accuracy is used to launch the adversarial attack (\Circled{\scriptsize\textbf{8}} in Fig.~\ref{fig:Toolflow}).

\textbf{Evaluation Metric:} To evaluate the performance of NeuroUnlock and DNN obfuscation, we use the LER metric to compute the difference between the recovered layer sequence and the original layer sequence, similarly to DeepSniffer~\cite{DeepSniffer} and NeurObfuscator~\cite{li2021neurobfuscator}. 
The LER is calculated as follows:
\begin{equation}
LER=\frac{ED(L,L^{*})}{|L^{*}|}
\end{equation}
where $L$ represents the predicted sequence, $L^{*}$ represents the ground-truth, and $|.|$ denotes the length of a sequence. $ED(p, q)$ denotes the edit distance between the $p$ and $q$ sequences, i.e., the minimum number of insertions, substitutions and deletions required to change $p$ into $q$ (also refered to as the \textit{Levenshtein} distance~\cite{LER}).

\section{Results}
\label{Results}
In this section, we demonstrate the effectiveness and performance of the NeuroUnlock and ReDLock techniques. We also perform a comparison between ReDLock and the state-of-the-art DNN obfuscation method.

\subsection{Effectiveness of NeuroUnlock and ReDLock}
\subsubsection{Random DNNs (\textit{dataset~B})}
\begin{figure}[!t]
     \centering
     \includegraphics[width=\linewidth]{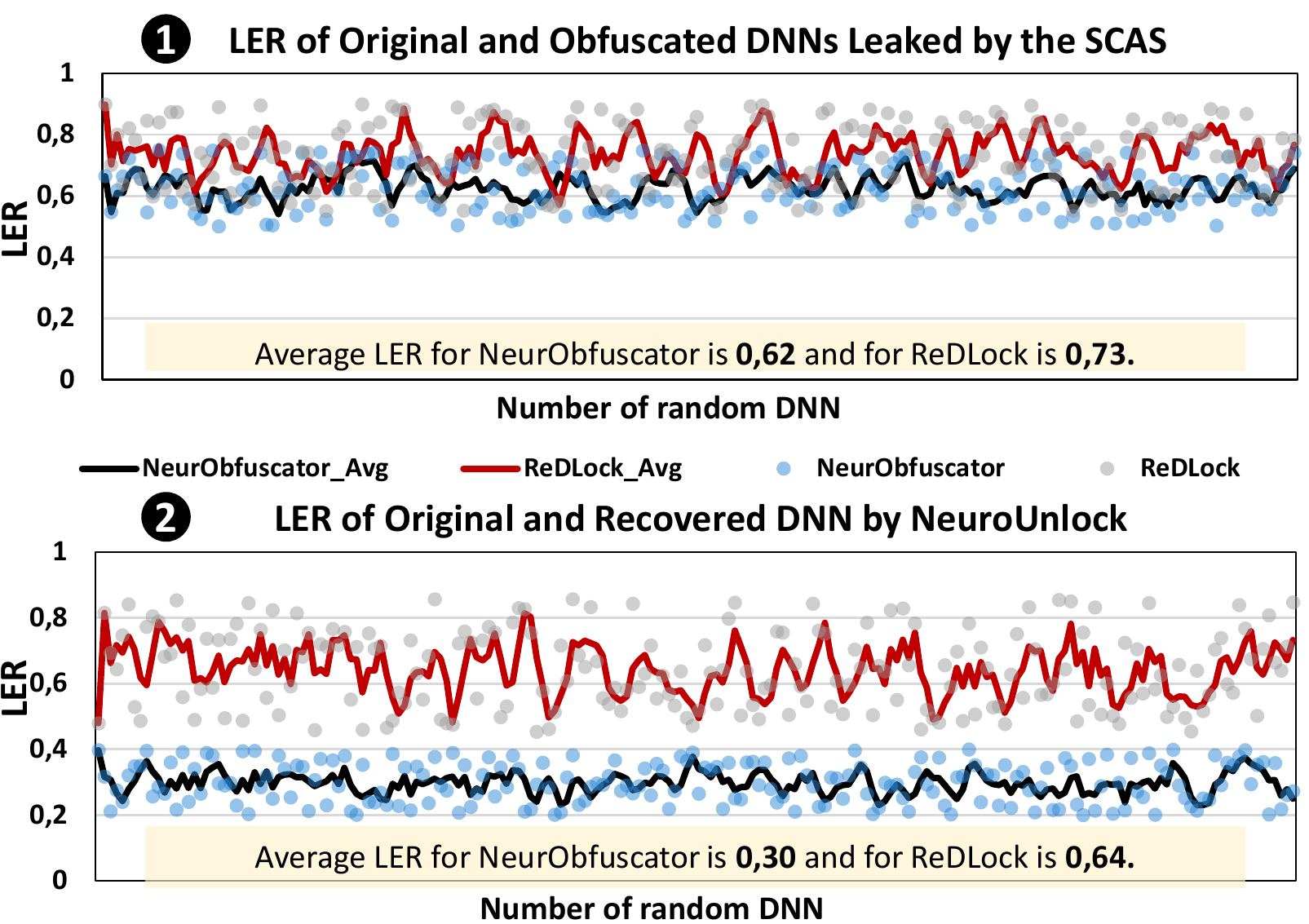}
     \caption{The SCAS attack and the NeuroUnlock LER results on $200$ random DNNs obfuscated using NeurObfuscator and ReDLock.
     }
     \vspace{-1em}
     \label{fig:random_LER}
\end{figure}

We launch NeuroUnlock on the $20\%$ of the DNNs in \textit{dataset~B} ($200$ DNNs), kept for testing and obfuscated using NeurObfuscator and ReDLock, attacking $200$ for each scenario, so $400$ DNNs in total. The results are reported in Fig.~\ref{fig:random_LER}. In~\Circled{\scriptsize\textbf{1}}, the LER between the original and the obfuscated DNNs recovered by the SCAS attack is presented (\textit{``Original-Obfuscated''}, $LER$~$1$ in Fig.~\ref{fig:Methodology}). In this dataset, the average LER for NeurObfuscator is $0.62$ and for ReDLock is $0.74$, which indicates a higher obfuscation level offered by ReDLock. It also shows that the SCAS attack cannot recover the original DNN when DNN obfuscation is in place. In~\Circled{\scriptsize\textbf{2}}, the LER between the original and the DNNs recovered by NeuroUnlock is presented (\textit{``Original-Recovered''}, $LER$~$2$ in Fig.~\ref{fig:Methodology}). The average LER obtained by NeuroUnlock when attacking NeurObfuscator is $0.2$, while it is $0.6$ when attacking ReDLock. These results show that NeuroUnlock can undo the na\"ive obfuscation, dropping the LER from an average of $0.62$ to an average of $0.2$, i.e., $67.7\%$ reduction (almost recovering the original architecture). The results also demonstrate that ReDLock is resilient to the proposed NeuroUnlock attack, achieving $2.16\times$ more resilience.
\subsubsection{Typical DNNs \textit{dataset~C}}
\begin{figure}[!t]
     \centering
     \includegraphics[width=\linewidth]{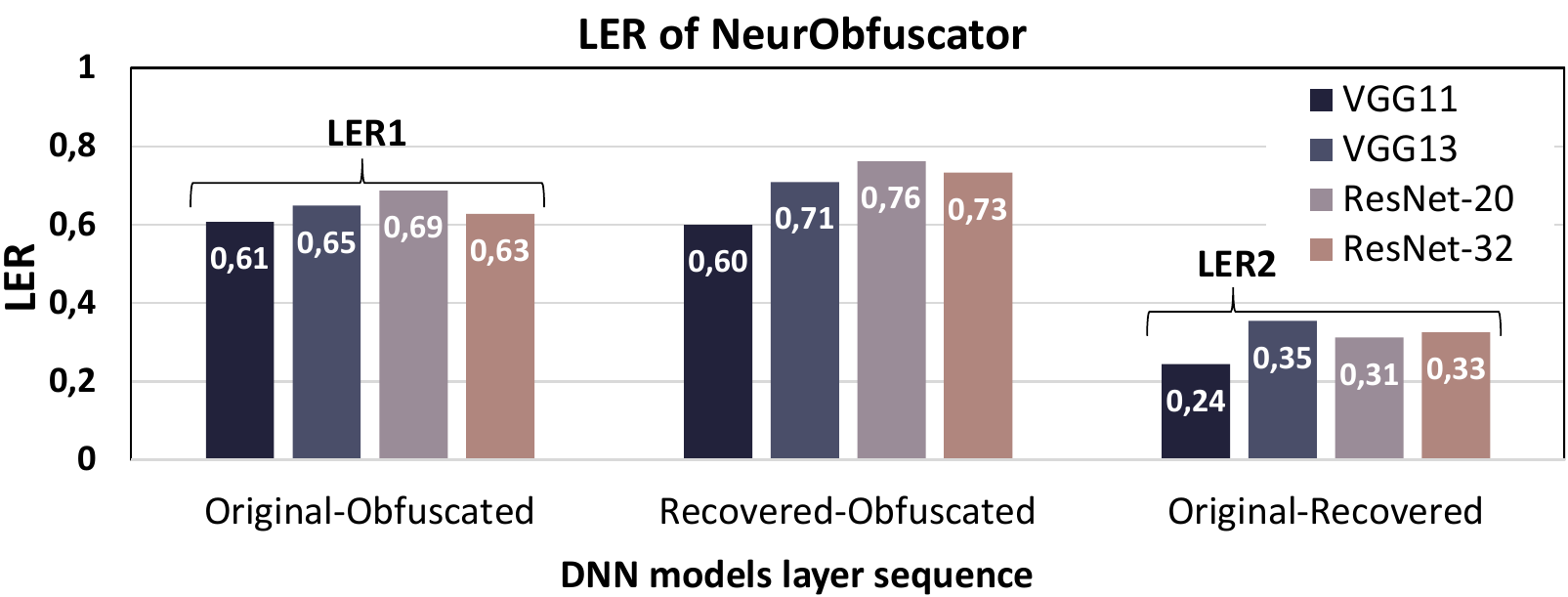}
     \caption{LER of NeuroUnlock and the SCAS attack on the DNNs obfuscated using NeurObfuscator. 
     }
     \vspace{-1em}
     \label{fig:LER_GA}
\end{figure}

Next, we launch NeuroUnlock and the SCAS attack discussed in Sec.~\ref{sec:SCAS_training} on the obfuscated DNNs in \textit{dataset~C} using NeurObfuscator, and report the LER of both attacks in Fig.~\ref{fig:LER_GA}.
We expect that the LER achieved by NeuroUnlock ($LER$~$2$) to be smaller than one reported by the SCAS attack ($LER$~$1$) since NeuroUnlock reverts the obfuscation. Indeed, the average $LER$~$1$ is $0.64$ while the average $LER$~$2$ is $0.31$. By launching NeuroUnlock after the SCAS attack, the attack LER drops by $52\%$ on average. We also measure the LER between the obfuscated DNN recovered by the SCAS attack and the ones recovered by NeuroUnlock (\textit{``Recovered-Obfuscated''} LER bar plots in Fig.~\ref{fig:LER_GA}). We expect a large LER value here to support that the DNNs recovered by NeuroUnlock are more similar to the original versions than they are to the obfuscated versions. Indeed, the results show that the \textit{``Recovered-Obfuscated''} LER is around $0.7$, compared to the average of $LER$~$2$ that is $0.31$ which means the original and recovered DNNs are more similar than recovered and obfuscated DNNs.

\begin{figure}[!t]
     \centering
     \includegraphics[width=\linewidth]{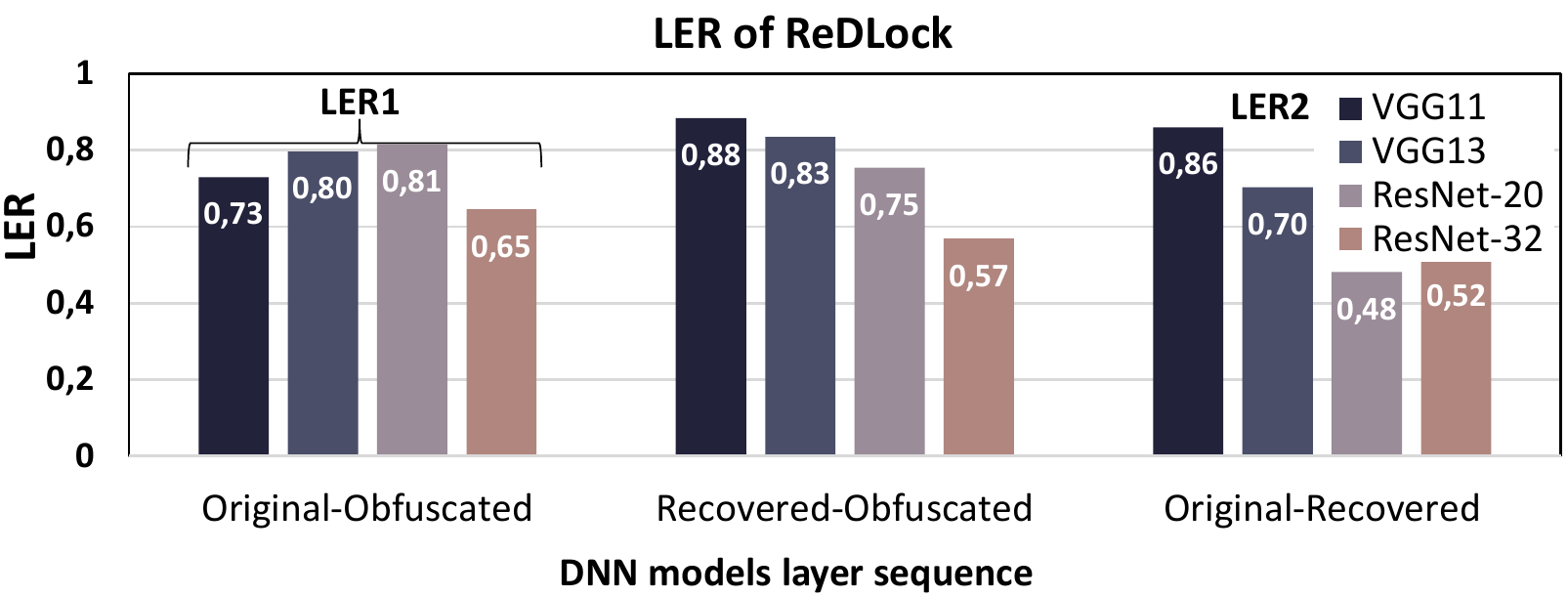}
     \vspace{-0.5em}
     \caption{LER of NeuroUnlock and the SCAS attack on the DNNs obfuscated using ReDLock.}
     \vspace{-1em}
     \label{fig:LER_rand}
\end{figure}

We repeat the same analysis (i.e., launch the SCAS attack and NeuroUnlock) against ReDLock, and report the results in Fig.~\ref{fig:LER_rand}. The average $LER$~$1$ and $LER$~$2$ are $0.74$ and $0.64$, respectively. By comparing the average of the differences of $LER$~$2$ in NeurObfuscator and RedLock, we demonstrate that ReDLock is $46$\% (2.16x) more resilient to NeuroUnlock than NeurObfuscator is.

\vspace{-0.5em}
\subsection{NeuroUnlock and ReDLock Performance Cost} 

\begin{figure}[!t]
     \centering
     \includegraphics[width=\linewidth]{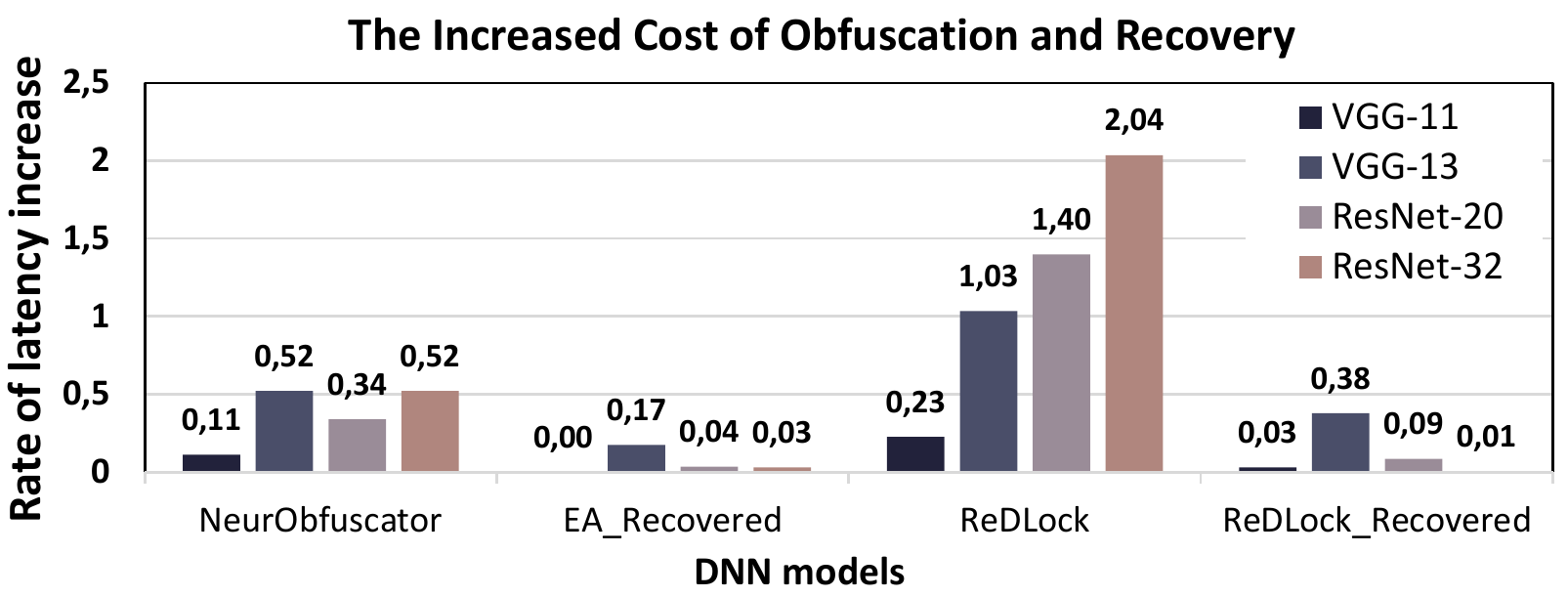}

     \vspace{-0.5em}
     \caption{The increase of latency cost in comparison with the baseline of original DNN for different models. The latency is calculated as the total number of GPU clock cycles to execute the DNN in inference mode.}
     \vspace{-1em}
     \label{fig:Latency}
\end{figure}

We compare the performance cost for the DNN models in \textit{dataset~C}. As shown in Fig. \ref{fig:Latency}, the obfuscation of DNNs using NeurObfuscator increases the latency of the original DNNs by $1.37\times$ on average, while each corresponding DNN, recovered by NeuroUnlock, shows very similar performance to the original DNN with only $1.06\times$ increase in latency.
Additionally, we report the latency of the obfuscated DNNs using ReDLock, showing that our obfuscation has merely $2\times$ latency increase compared to the original DNNs, which is slightly higher than the latency cost of NeurObfuscator.
In NeurObfuscator, the final obfuscation is selected based on the user budget for the latency, so it is optimized for performance. On the other hand, the ReDLock obfuscation scheme is randomized and only considers the LER of the obfuscated DNNs without considering the latency cost. Nevertheless, the latency cost is still in the acceptable range.

\begin{figure}[!t]
     \centering
     \includegraphics[width=\linewidth]{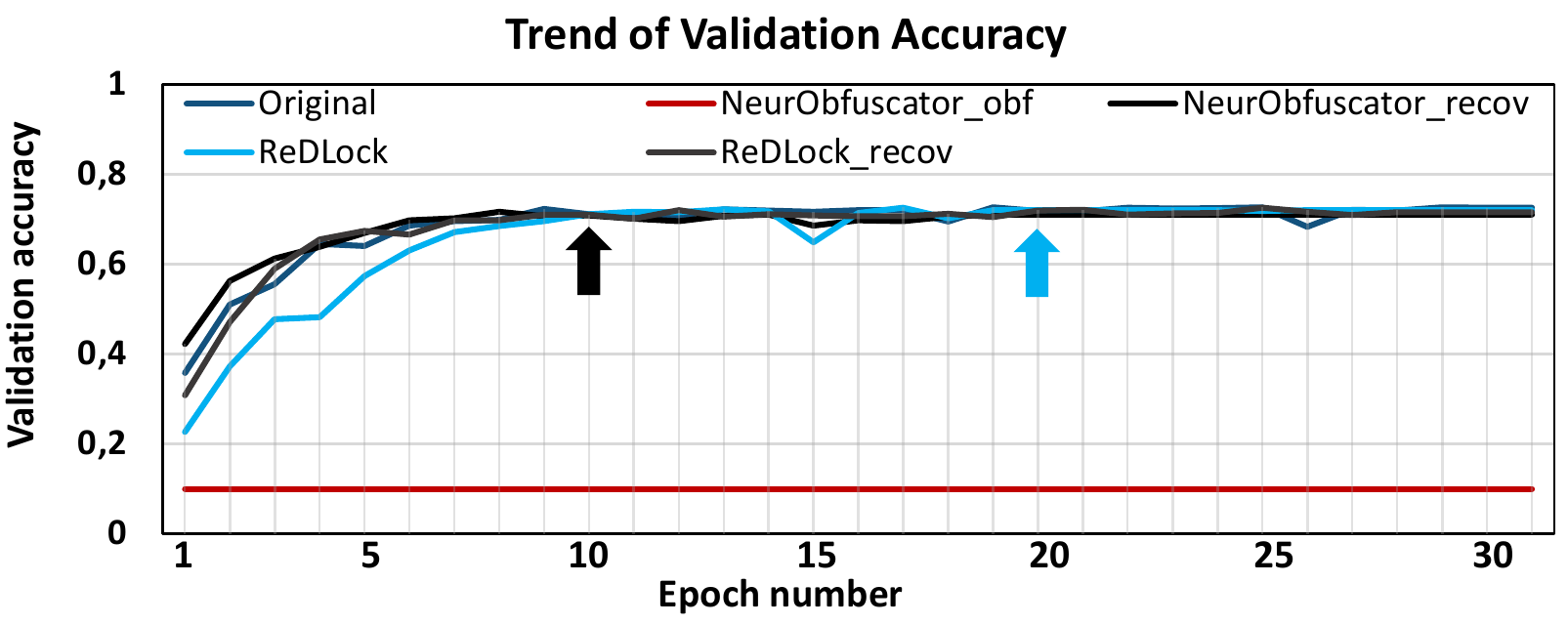}
   
     \vspace{-1em}
     \caption{Validation accuracy after training all models. In this Figure, the convergence point of obfuscated DNN using ReDLock is marked by the blue arrow, and the original and recovered models are converged at epoch 10, marked by the black arrow. The DNN obfuscated using NeurObfuscator does not converge (red).
     }
     \vspace{-1em}
     \label{fig:Training}
\end{figure}

Next, we evaluate the training results for the original, obfuscated, and recovered DNNs by NeuroUnlock, for NeurObfuscator and ReDLock. We compare these models for validation accuracy for $30$ epochs of training with CIFAR-10 dataset~\cite{cifar10Krizhevsky2009LearningML} on VGG-11 DNN~\cite{vgg}. As shown in Fig.~\ref{fig:Training}, original, and recovered DNNs show similar validation accuracy trends (maximum accuracy of $70\%$). The results show the training accuracy of recovered DNN is only $1.4\%$ less than the original DNN, while the ReDLock Obfuscated DNN converges slower. In our experiments, the obfuscated DNNs by NeurObfuscator did not converge in training and the validation accuracy stayed close to $10\%$ for $30$ epochs which is comparable to random classification. 
\vspace{-0.5em}
\subsection{Adversarial Attack Results}
 To perform the adversarial attack, first, we build a training dataset, starting from the CIFAR-10 images and using the label output of the target model (VGG-11). The success rate of the adversarial attack is defined as the number of successfully misclassified adversarial samples with respect to the total number of tested input samples. Next, we analyze the adversarial attack results in three scenarios using the extracted architectures to generate adversarial samples (see Fig.~\ref{fig:Adv_result}).

\subsubsection{Original Model} To test the attack setup and show its effectiveness, we utilize the original model to generate the adversarial samples (\Circled{\scriptsize\textbf{1}} in Fig.~\ref{fig:Adv_result}). The average of success rate for the adversarial attack in the original DNN is $98\%$. 
\subsubsection{Public Family DNNs}	We generate the adversarial samples using publicly available DNNs of different families, such as GoogleNet (\Circled{\scriptsize\textbf{2}} in Fig.~\ref{fig:Adv_result}), Inception-V3 (\Circled{\scriptsize\textbf{3}} in Fig.~\ref{fig:Adv_result}) and ResNet-34 (\Circled{\scriptsize\textbf{4}} in Fig.~\ref{fig:Adv_result}), implemented in PyTorch~\cite{cifar10pytorchmodels}. The results show an average success rate of $14\%$, $48\%$, and $88\%$, respectively.
\subsubsection{Obfuscated and Recovered} 
\begin{figure}[!t]
     \centering
     \includegraphics[width=\linewidth]{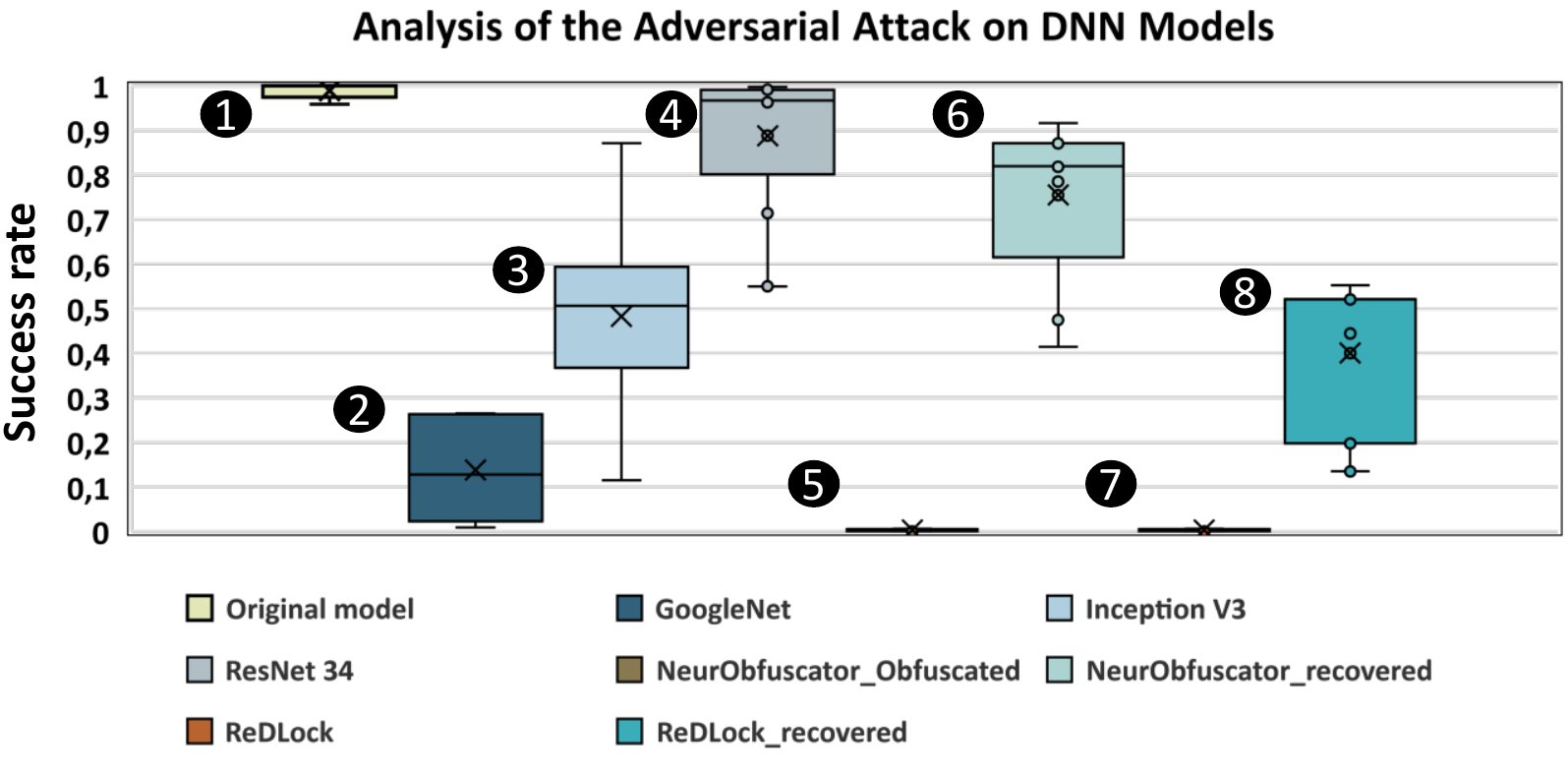}
          \vspace{-1em}
     \caption{Comparison of the average success rate for VGG-11 as the original black-box model, obfuscated using NeurObfuscator and ReDLock. Additionally, the average success rate for various other DNN families.}
     \vspace{-1em}
     \label{fig:Adv_result}
\end{figure}

Next, we use the obfuscated DNN using the NeurObfuscator technique to generate the samples. The results show that the attack was unsuccessful with an average success rate of $0.3\%$ (\Circled{\scriptsize\textbf{5}} in Fig.~\ref{fig:Adv_result}). Our experiment shows that applying NeuroUnlock on extracted obfuscated DNN, increases the success rate to $75\%$ (\Circled{\scriptsize\textbf{6}} in Fig.~\ref{fig:Adv_result}) with the average of $51.7\%$. Also, the experiments indicate that ReDLock-obfuscated DNN has $0.3\%$ success rate (\Circled{\scriptsize\textbf{7}} in Fig.~\ref{fig:Adv_result}) while the recovered DNN has $40\%$ average in success rate (\Circled{\scriptsize\textbf{8}} in Fig.~\ref{fig:Adv_result}). These results indicate that stronger DNN obfuscation can mitigate the adversarial attacks, and NeuroUnlock-recovered DNN improves the effectiveness of the adversarial attack.


\subsection{Future Works}
This work exposes the vulnerability of state-of-the-art DNN obfuscation methods (based on predictable and reversible modifications employed in a given DNN architecture) to side-channel-based architecture stealing. In future work directions, we plan to extend experiments of NeuroUnlock attack for the DNNs with complex architectures, e.g., transformers and various datasets, i.e., CIFAR100 and ImageNet.
Furthermore, we will explore how to extend the ReDLock methodology to increase the resilience of the obfuscation algorithm while maintaining the latency and performance of original DNNs.

\section{Conclusion}
\label{Conclusion}
In this paper, we have proposed \textit{NeuroUnlock} as an efficient side-channel-based method for extracting the architecture of obfuscated deep neural networks (DNNs), revealing the vulnerabilities of the existing DNN obfuscation methods (i.e., predictable and reversible modifications in the architecture of a given DNN). NeuroUnlock adopts a machine learning (ML) algorithm that learns the obfuscation procedure and reverts it. The recovered architecture is then used to build a duplicate model and launch a subsequent adversarial attack. Our results show that NeuroUnlock improves the success rate of the adversarial attack to $75$\%. After exposing the vulnerabilities of existing DNN obfuscation methods, we propose \textit{ReDLock} as a novel defense solution against NeuroUnlock and other side-channel-based architecture stealing (SCAS) attacks. ReDLock obfuscates the architecture of the DNN following a random obfuscation approach while maintaining the original functionality of the network. ReDLock does not have an obfuscation pattern and consequently cannot be learned by ML algorithms. We believe that our work opens new doors for future investigations of DNN obfuscation.
\section{Acknowledgement}
This work has been supported by the Doctoral College Resilient
Embedded Systems, which is run jointly by the TU Wien's Faculty of
Informatics and the UAS Technikum Wien.
It was also jointly supported by the NYUAD Center for Interacting Urban Networks (CITIES), funded by Tamkeen under the NYUAD Research Institute Award CG001 and Center for CyberSecurity (CCS), funded by Tamkeen under the NYUAD Research Institute Award G1104.

\bibliographystyle{IEEEtran}
\bibliography{main}

\begin{thebibliography}{10}
\providecommand{\url}[1]{#1}
\csname url@samestyle\endcsname
\providecommand{\newblock}{\relax}
\providecommand{\bibinfo}[2]{#2}
\providecommand{\BIBentrySTDinterwordspacing}{\spaceskip=0pt\relax}
\providecommand{\BIBentryALTinterwordstretchfactor}{4}
\providecommand{\BIBentryALTinterwordspacing}{\spaceskip=\fontdimen2\font plus
\BIBentryALTinterwordstretchfactor\fontdimen3\font minus
  \fontdimen4\font\relax}
\providecommand{\BIBforeignlanguage}[2]{{%
\expandafter\ifx\csname l@#1\endcsname\relax
\typeout{** WARNING: IEEEtran.bst: No hyphenation pattern has been}%
\typeout{** loaded for the language `#1'. Using the pattern for}%
\typeout{** the default language instead.}%
\else
\language=\csname l@#1\endcsname
\fi
#2}}
\providecommand{\BIBdecl}{\relax}
\BIBdecl

\bibitem{kato2015open}
S.~Kato, E.~Takeuchi, Y.~Ishiguro, Y.~Ninomiya, K.~Takeda, and T.~Hamada, ``An
  open approach to autonomous vehicles,'' \emph{IEEE Micro}, vol.~35, no.~6,
  pp. 60--68, 2015.

\bibitem{tesla}
\BIBentryALTinterwordspacing
``Autopilot.'' [Online]. Available: \url{https://www.tesla.com/autopilot}
\BIBentrySTDinterwordspacing

\bibitem{krizhevsky2012imagenet}
A.~Krizhevsky, I.~Sutskever, and G.~E. Hinton, ``{ImageNet} classification with
  deep convolutional neural networks,'' \emph{Advances in Neural Information
  Processing Systems (NIPS)}, vol.~25, pp. 1097--1105, 2012.

\bibitem{chakrabortySurveyAdversarialAttacks2021}
A.~Chakraborty, M.~Alam, V.~Dey, A.~Chattopadhyay, and D.~Mukhopadhyay, ``A
  survey on adversarial attacks and defences,'' \emph{CAAI Transactions on
  Intelligence Technology}, vol.~6, pp. 25--45, 03 2021.

\bibitem{DeepSniffer}
X.~Hu, L.~Liang, S.~Li, L.~Deng, P.~Zuo, Y.~Ji, X.~Xie, Y.~Ding, C.~Liu,
  T.~Sherwood, and Y.~Xie, ``{DeepSniffer}: A {DNN} model extraction framework
  based on learning architectural hints,'' in \emph{Proceedings of the
  Twenty-Fifth International Conference on Architectural Support for
  Programming Languages and Operating Systems (ASPLOS)}, 2020, p. 385–399.

\bibitem{goldreich1996software}
O.~Goldreich and R.~Ostrovsky, ``Software protection and simulation on
  oblivious {RAMs},'' \emph{Journal of the ACM (JACM)}, vol.~43, no.~3, pp.
  431--473, 1996.

\bibitem{shi2011oblivious}
E.~Shi, T.-H.~H. Chan, E.~Stefanov, and M.~Li, ``Oblivious {RAM} with o
  (({logN}) 3) worst-case cost,'' in \emph{Advances in Cryptology
  (ASIACRYPT)}.\hskip 1em plus 0.5em minus 0.4em\relax Springer, 2011, pp.
  197--214.

\bibitem{karimi2020hardware}
E.~Karimi, Y.~Fei, and D.~Kaeli, ``Hardware/software obfuscation against timing
  side-channel attack on a {GPU},'' in \emph{International Symposium on
  Hardware Oriented Security and Trust (HOST)}.\hskip 1em plus 0.5em minus
  0.4em\relax IEEE, 2020, pp. 122--131.

\bibitem{li2021neurobfuscator}
J.~Li, Z.~He, A.~S. Rakin, D.~Fan, and C.~Chakrabarti, ``{NeurObfuscator}: A
  full-stack obfuscation tool to mitigate neural architecture stealing,'' in
  \emph{International Symposium on Hardware Oriented Security and Trust
  (HOST)}.\hskip 1em plus 0.5em minus 0.4em\relax IEEE, 2021.

\bibitem{hwsecurity}
Q.~Xu, M.~T. Arafin, and G.~Qu, ``Security of neural networks from hardware
  perspective: A survey and beyond,'' in \emph{Asia and South Pacific Design
  Automation Conference (ASP-DAC)}.\hskip 1em plus 0.5em minus 0.4em\relax
  IEEE, 2021, pp. 449--454.

\bibitem{liu2015ghostrider}
C.~Liu, A.~Harris, M.~Maas, M.~W. Hicks, M.~Tiwari, and E.~Shi, ``Ghostrider:
  {A} hardware-software system for memory trace oblivious computation,'' in
  \emph{Proceedings of the Twentieth International Conference on Architectural
  Support for Programming Languages and Operating Systems, (ASPLOS)}, 2015, pp.
  87--101.

\bibitem{jagielski2020high}
M.~Jagielski, N.~Carlini, D.~Berthelot, A.~Kurakin, and N.~Papernot, ``High
  accuracy and high fidelity extraction of neural networks,'' in \emph{USENIX
  Security Symposium}, 2020, pp. 1345--1362.

\bibitem{rakin2021deepsteal}
A.~S. Rakin, M.~H.~I. Chowdhuryy, F.~Yao, and D.~Fan, ``{DeepSteal}: Advanced
  model extractions leveraging efficient weight stealing in memories,'' in
  \emph{IEEE Symposium on Security and Privacy (S\&P)}, 2021.

\bibitem{lstm}
\BIBentryALTinterwordspacing
S.~Hochreiter and J.~Schmidhuber, ``\BIBforeignlanguage{en}{Long {Short}-{Term}
  {Memory}},'' \emph{\BIBforeignlanguage{en}{Neural Computation}}, vol.~9,
  no.~8, pp. 1735--1780, Nov. 1997. [Online]. Available:
  \url{https://direct.mit.edu/neco/article/9/8/1735-1780/6109}
\BIBentrySTDinterwordspacing

\bibitem{zhu2019eena}
H.~Zhu, Z.~An, C.~Yang, K.~Xu, E.~Zhao, and Y.~Xu, ``{EENA}: efficient
  evolution of neural architecture,'' in \emph{2019 IEEE/CVF International
  Conference on Computer Vision Workshop (ICCVW)}, 2019, pp. 1891--1899.

\bibitem{chen2018tvm}
T.~Chen, T.~Moreau, Z.~Jiang, L.~Zheng, E.~Yan, H.~Shen, M.~Cowan, L.~Wang,
  Y.~Hu, L.~Ceze \emph{et~al.}, ``{TVM}: An automated end-to-end optimizing
  compiler for deep learning,'' in \emph{USENIX Security Symposium}, 2018, pp.
  578--594.

\bibitem{dataset}
E.~D. Cubuk, B.~Zoph, J.~Shlens, and Q.~Le, ``{RandAugment}: Practical
  automated data augmentation with a reduced search space,'' in \emph{Advances
  in Neural Information Processing Systems (NIPS)}, H.~Larochelle, M.~Ranzato,
  R.~Hadsell, M.~F. Balcan, and H.~Lin, Eds., vol.~33, 2020, pp.
  18\,613--18\,624.

\bibitem{lecunBackpropagationAppliedHandwritten1989}
\BIBentryALTinterwordspacing
Y.~LeCun, B.~Boser, J.~S. Denker, D.~Henderson, R.~E. Howard, W.~Hubbard, and
  L.~D. Jackel, ``Backpropagation {{Applied}} to {{Handwritten Zip Code
  Recognition}},'' vol.~1, no.~4, pp. 541--551. [Online]. Available:
  \url{10/bknd8g}
\BIBentrySTDinterwordspacing

\bibitem{Nsight}
Nvidia nsight systems: {\url{https://developer.nvidia.com/nsight-systems}}.
  Accessed: 2022-02-10.

\bibitem{OpenNMT}
G.~Klein, Y.~Kim, Y.~Deng, J.~Senellart, and A.~Rush, ``{O}pen{NMT}:
  Open-source toolkit for neural machine translation,'' in \emph{Proceedings of
  {ACL} 2017, System Demonstrations}.\hskip 1em plus 0.5em minus 0.4em\relax
  Association for Computational Linguistics, Jul. 2017, pp. 67--72.

\bibitem{AdvAtt}
S.~Cheng, Y.~Dong, T.~Pang, H.~Su, and J.~Zhu, ``Improving black-box
  adversarial attacks with a transfer-based prior,'' in \emph{Proceedings of
  the 33rd International Conference on Neural Information Processing Systems
  (NIPS)}.\hskip 1em plus 0.5em minus 0.4em\relax Curran Associates Inc., 2019.

\bibitem{pytorch}
A.~Paszke, S.~Gross, F.~Massa, A.~Lerer, J.~Bradbury, G.~Chanan, T.~Killeen,
  Z.~Lin, N.~Gimelshein, L.~Antiga, A.~Desmaison, A.~Kopf, E.~Yang, Z.~DeVito,
  M.~Raison, A.~Tejani, S.~Chilamkurthy, B.~Steiner, L.~Fang, J.~Bai, and
  S.~Chintala, ``Pytorch: An imperative style, high-performance deep learning
  library,'' in \emph{Advances in Neural Information Processing Systems
  (NIPS)}.\hskip 1em plus 0.5em minus 0.4em\relax Curran Associates, Inc.,
  2019, pp. 8024--8035.

\bibitem{sgd}
H.~Robbins and S.~Monro, ``A {Stochastic} {Approximation} {Method},'' \emph{The
  Annals of Mathematical Statistics}, vol.~22, no.~3, pp. 400--407, Sep. 1951.

\bibitem{cifar10Krizhevsky2009LearningML}
A.~Krizhevsky, ``Learning multiple layers of features from tiny images,'' 2009.

\bibitem{DBLP:journals/corr/KingmaB14}
D.~P. Kingma and J.~Ba, ``Adam: {A} method for stochastic optimization,'' in
  \emph{International Conference on Learning Representations (ICLR)}, Y.~Bengio
  and Y.~LeCun, Eds., 2015.

\bibitem{LER}
G.~Navarro, ``A guided tour to approximate string matching,'' \emph{ACM Comput.
  Surv.}, vol.~33, no.~1, p. 31–88, mar 2001.

\bibitem{vgg}
K.~Simonyan and A.~Zisserman, ``Very deep convolutional networks for
  large-scale image recognition,'' in \emph{International Conference on
  Learning Representations (ICLR)}, Y.~Bengio and Y.~LeCun, Eds., 2015.

\bibitem{cifar10pytorchmodels}
\BIBentryALTinterwordspacing
``huyvnphan/{PyTorch}\_cifar10,'' Jan. 2021. [Online]. Available:
  \url{https://zenodo.org/record/4431043}
\BIBentrySTDinterwordspacing

\end{thebibliography}

\end{document}